\documentclass[%
preprint,
 amsmath,amssymb,
 aps,
]{revtex4-2}

\usepackage{graphicx}
\usepackage{dcolumn}
\usepackage{bm}

\begin{document}

\title{Drop Impact Dynamics of Complex Fluids: A Review}

\author{Phalguni Shah}
\author{Michelle M.\ Driscoll}
\affiliation{%
 Department of Physics \& Astronomy, Northwestern University
}%

\begin{abstract}
The impact of fluid drops on solid substrates has widespread interest in many industrial coating and spraying applications, such as ink-jet printing and agricultural pesticide sprays. Many of the fluids used in these applications are non-Newtonian, that is they contain particulate or polymeric additives that strongly modify their flow behaviour.
While a large body of experimental and theoretical work has been done to understand the impact dynamics of Newtonian fluids, we as a community have much progress to make to understand how these dynamics are modified when the impact fluid has non-Newtonian rheology. In this review, we outline recent experimental, theoretical, and computational 
advances in the study of impact dynamics of complex fluids on solid surfaces. Here, we provide an overview of this field that is geared towards a multidisciplinary audience. Our discussion is segmented by two principal material constitutions: polymeric fluids and particulate suspensions.  Throughout, we highlight promising future directions, as well as ongoing experimental and theoretical challenges in the field.
\end{abstract}

\maketitle


\newpage

Newtonian fluids, such as water, have a viscosity that does not change with applied shear; that is, their flow behaviour is independent of forcing. This is not true for non-Newtonian fluids: they exhibit strikingly different flow behaviours as a function of applied stress, such as a large reduction in a viscosity or  even a solid-like response. Such complex fluids are ubiquitous in everyday life; some examples are  paint, blood, ketchup, and shampoo. To understand how to process and manipulate these materials requires characterization of their flow behaviours, with the aim of elucidating a constitutive relation that links the fluid microstructure to its macroscopic material properties. 
Rheometry is traditionally used to characterise the flow of complex fluids, and it has proved invaluable in providing a quantitative understanding of bulk non-Newtonian flow behaviours. In brief, rheometric  techniques enable one to apply a known shear rate or stress and then quantify the material response. However, accessing regimes of very high stresses or obtaining localised flow information is challenging with these tools; new techniques are required to probe the underlying physics that drives localised deformations under at high stresses and transient flows.

Recent work has established drop impact as a valuable system for addressing this challenge. These studies allow us to explore material behaviour at high and dynamically changing shear. For example, a fluid drop of diameter $d_0 = 3$ mm impacting a surface at a velocity of $u_0 = 5$ m/s undergoes an \emph{instantaneous and localised} strain rate of $\dot{\gamma}=\frac{u_0}{d_0}=1.67\times 10^3$ $ s^{-1}$ at impact; this is approaching the operating limit of many rheometers. In case of impacting complex fluids, this high and instantaneous strain rate  gives rise to a wide variety of phenomena that are quite distinct and not observed in the impact of Newtonian fluids. The deformable free surface in an impacting drop allows one to directly observe the material's response as the shear rate varies over space and time. Drop impact experiments therefore provide a bridge to connect localised manifestations of complex fluid behaviours to bulk rheological measurements and constitutive models. When applied to complex fluids, drop impact can uncover important localised details and compliment bulk rheological work to inform detailed constitutive models of multi-phase fluids. In this review, we summarize key insights from impact studies of complex fluids. Along the way, we attempt to draw broad connections with both complex fluid rheology and Newtonian drop impact studies.
 
\begin{figure*}[h]
\centering
\includegraphics[height = 5.5cm]{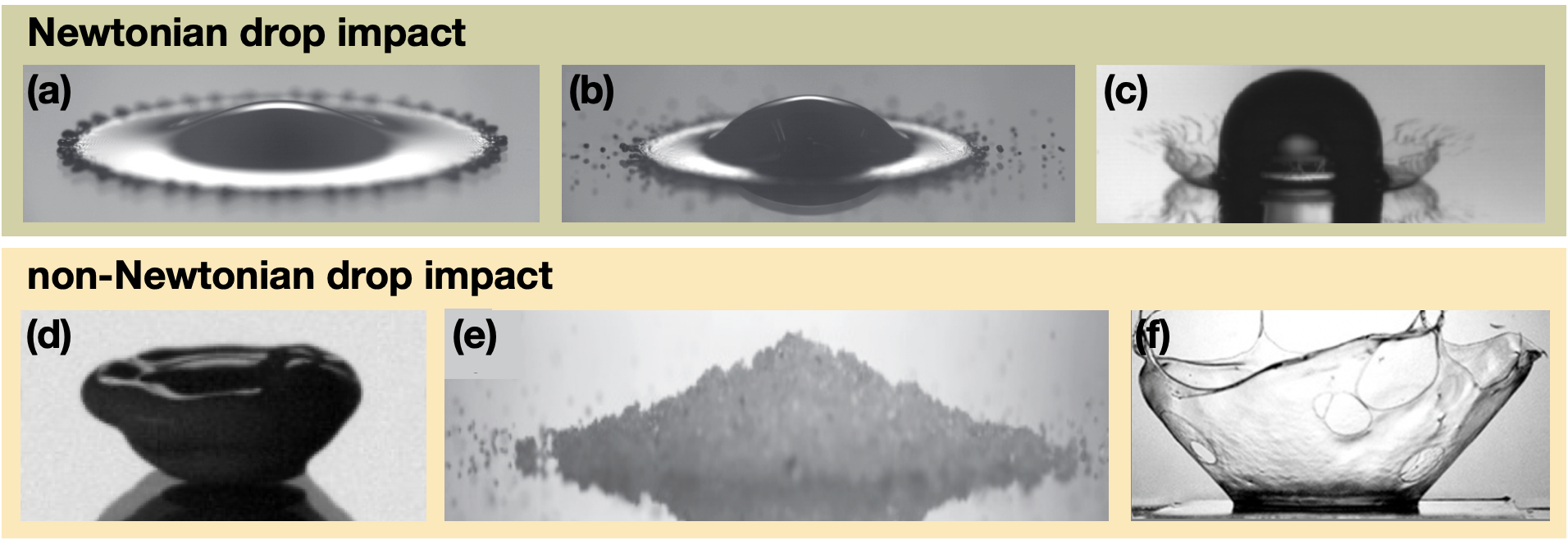}
    \caption  {\textbf{behaviour of impacting drops of Newtonian (a-c) and complex fluids (d-f).} (a) A drop of mercury (a Newtonian fluid) spreading on glass (Adapted from~\cite{JosserandReview}), (b) a drop of mercury undergoing a prompt splash when impacted on superhydrophobized glass (Adapted from~\cite{JosserandReview}), (c) a drop of silicone oil exhibiting corona splash on glass, (d) a drop of silica in water ($\phi=49 \%$) partially shear jammed after impacting on glass (Adapted from~\cite{Shah_2021_impact}), (e) a drop of dense granular suspension after impacting on glass (Adapted from~\cite{Marston}), and (f) A drop of 0.25\% Carbopol in water impacting  a wet substrate (Adapted from~\cite{Blackwell2015}).
}
    \label{fig:intro}
\end{figure*}

Drop impact is a challenging fluid dynamic problem, as the relevant parameter space is vast.  In addition to the impact conditions (e.g.\ impact velocity, drop size, impact angle), the dynamics of drop impact is controlled by  the bulk properties of the fluid (e.g.\ density, surface tension, viscosity), the properties of the substrate (e.g.\ wettability, surface roughness), and those of the surrounding gas (e.g.\ density, pressure, and molecular weight)~\cite{YarinReview, JosserandReview}. As a community, we have been working to untangle the complex and nuanced physics that governs drop impact for over a century ~\cite{Worthington, YarinReview, JosserandReview}. In addition to the spreading dynamics, a rich variety of phenomena and subtleties have been uncovered for Newtonian fluids, such as numerous distinct splashing mechanisms~\cite{YarinReview} and a pressure-controlled splashing threshold~\cite{Xu2005}. When impacted, complex fluid drops can exhibit an even more diverse array of behaviours such as solidification and delayed spreading upon impact \textbf{[Fig.\ ~\ref{fig:intro}]}. Drops of complex fluids impacting on a substrate are ubiquitous in industrial and manufacturing processes such as spraying and coating of surfaces~\cite{breitenbach2018drop}, inkjet printing~\cite{lohse2022fundamental}, agrochemical delivery, forensics~\cite{smith2018wetting}, and pharmaceutical manufacturing~\cite{bolleddula2010impact}. While some
impact behaviours can be understood using Newtonian drop impact as a model, some other regimes exhibit drastically different phenomena with rich underlying mechanisms. A more comprehensive understanding of impacting complex fluid drops is paramount for controlling a range of industrial processes.

There have been a number of studies on the impact of complex fluids focused on specific regions of the parameter space to aid in the design of functional materials.  However, even more so than Newtonian fluid impact, the parameter space is vast, and most studies have been highly focused, for example exploring a set of questions pertaining to a specific application. Bertola and Marengo~\cite{bertola2012bookchapter} have provided a useful review on impacting non-Newtonian drops, mainly focused on polymeric fluids. Recent work by Aksoy et al.~\cite{AKSOY2022nanofluid} discusses the effect of nanoparticle additives on the splashing of fluid drops, and shows that nanoparticles have a significant effect on splashing even in the dilute regime. Although the flow properties of complex fluids are highly varied, the conditions in drop impact are identical to Newtonian drop impact studies and thus examining a broader range of fluids through the lens of Newtonian impact may help us address fundamental questions about complex fluid flow. The purpose of this review is to draw common themes among the disparate works on impacting complex fluids, and connect these observations to fluid rheology wherever possible. Being geared towards a multidisciplinary audience, we dedicate a section to discuss key rheological concepts for the uninitiated. We hope that this review may serve as a useful reference point to both fundamental and applied researchers investigating complex fluids. Furthermore, the connections drawn here may facilitate more systematic work in the future, so as to develop an understanding of complex flow under dynamic, free-surface conditions. For the purposes of this review, we classify past work in drop impact of complex fluids in two major categories: particulate suspensions and polymeric fluids. Much of the work summarized here is confined to impact on dry, solid surfaces. However, as impact on compliant or wet surfaces is broadly relevant in many processes such as spraying pesticides onto crop vegetation, we will discuss these aspects briefly.

The review is structured as follows:  Section \ref{concepts} discusses relevant rheological and fluid dynamic concepts that make an appearance throughout the review. Section \ref{Newtonian}  provides an overview of the current state of the Newtonian drop impact field, and Section \ref{methods} describes common experimental and computational approaches to the drop impact problem. We have organized the remaining sections by material composition.  In section \ref{polymeric}, we summarize both experimental and numerical results for the impact of polymeric drops. The advances in the impact dynamics of  particulate suspensions are discussed in section \ref{particulate}; we have divided these studies into impacts governed by bulk rheological behaviour, and impacts which are best characterised by considering particle inertia. Section \ref{conclusion} concludes the review by outlining the current challenges in the field, and suggests several directions of research that would lead to a unified understanding of the impact of complex fluids on solid substrates.

\section {Relevant concepts and parameters}\label{concepts}

In this section, we describe key properties of complex fluids relevant to drop impact. For a more thorough description of non-Newtonian flow and rheological properties, we refer the reader to the following sources~\cite{Wagner2012colloidal,shaw2012introduction,macosko1994rheology}.

\begin{figure*}[h]
\centering
\includegraphics[height = 9cm]{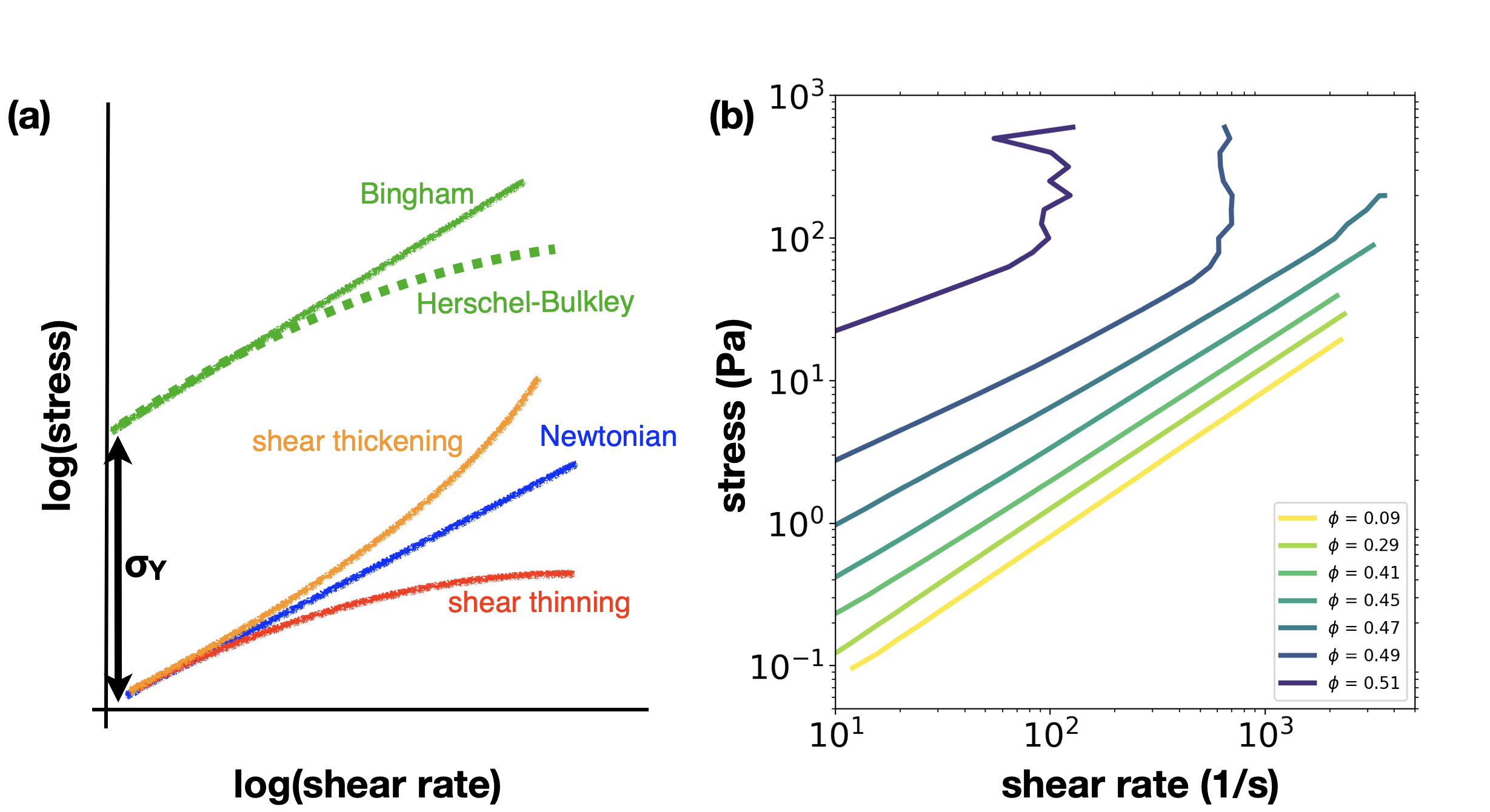}
    \caption  {\textbf{Rheology of complex fluids.} (a) Schematic illustrating flow properties of common complex fluids.  Newtonian liquids (blue) have a linear relationship between stress and shear rate, reflecting that their viscosity is constant (viscosity is defined as stress/shear rate).  Non-Newtonian fluids have a non-linear relation between stress and shear rate, the two common examples are fluids which shear thicken or shear thin.  Shear thinning fluids (red) have a viscosity which decreases with increasing shear rate, and thus a decreasing relationship between stress and shear rate. Shear thickening fluids (orange) have increase in viscosity with shear rate, and thus have an increasing relationship between stress and shear rate. Viscoplastic fluids (green) display a yield stress, $\sigma_Y$ (finite stress is required to create flow), and then either shear thin (Hershcel-Bulkley, dashed curve) or behave as Newtonian fluids (Bingham, solid curve).   (b) Rheological data from a dense colloidal suspension (830 nm silica spheres in water).  At low volume fractions, $\phi$, the suspension behaves as a more viscous Newtonian fluid.  As the volume fraction is increased, the fluid displays a yield stress and weak shear thinning, and then at the highest volume fractions, shear thickening behaviour appears.
}
    \label{fig:rheo}
\end{figure*}
  
In a Newtonian fluid, the viscosity is a constant material parameter that characterises the fluid's resistance to flow.  In contrast, the viscosity of complex fluid changes as a function of the applied shear, giving rise to exotic and unexpected behaviour such as rod-climbing in polymeric fluids, and transient solidification in dense suspensions~\cite{Wagner2012colloidal,DennMortonSoftMatter, osswald2015polymer}. Among an enormous variety of non-Newtonian responses, the most commonly observed are shear thinning and shear thickening [Fig.\ \ref{fig:rheo}a].  \emph{Shear thinning} is defined as a \emph{decrease} in the fluid viscosity as applied shear is increased. As its name implies, \emph{shear thickening} is the opposite, where the fluid viscosity \emph{increases} with shear. Furthermore, the same fluid can exhibit both shear-thinning and thickening depending on the applied stress; dense suspensions often shear thin at low shear stresses and then begin to shear thicken as the shear stress is increased. In each regime, the relationship between shear stress and shear rate can be described as a power-law:

\begin{equation}\label{powerlaw}
    \tau_{shear}=k\dot\gamma^n,
\end{equation}

so that $n=1$ corresponds to Newtonian flow profile, $n<1$ to shear thinning, and $n>1$ to shear thickening. 
Some complex fluids exhibit both elastic and viscous behaviour when they are deformed, known as viscoelasticity.  Many viscoelastic fluids have a yield stress, a critical stress, $\tau_Y$, above which the the fluid transitions from elastic-dominant(solid-like) to viscous-dominated (liquid-like) flow. Practically, this transition beyond the yield stress often manifests as a dramatic drop in the fluid viscosity~\cite{BARNES1999133,Balmforth2014, COUSSOT201431}. Many everyday fluids such as ketchup and toothpaste are yield-stress fluids; we are all familiar of the experience of applying a large shear stress (above $\tau_Y)$ to a bottle of ketchup or other thick sauce to induce flow. Many yield-stress fluids often also exhibit shear thinning at higher shear rates.  Although there are additional nuances to the measured flow behaviour of such fluids,  the concept of a critical yield stress is useful to model materials that exhibit this dramatic change in flow behaviour.  The Herschel-Bulkley model~\cite{Herschel1926} incorporates yield stress into equation \ref{powerlaw}:
\begin{equation} \label {Herschel}
    \tau=\tau_Y+k\dot\gamma^n \mbox{ for } \tau>\tau_Y,
\end{equation} 

where $\tau_Y$ is the yield stress of the fluid. 

 
The flow behaviour of complex fluids can also be highly dependent on the direction of applied shear. For example, in addition the shear viscosity, some complex fluids (typically polymeric fluids) exhibit a change in flow under elongational stresses, characterised by the elongational viscosity~\cite{extensional_PETRIE20061}. Elongational viscosity and viscoelasticity are more prevalent in polymeric fluids, and are thought to be connected to the geometric deformation of polymer chains~\cite{shaw2012introduction}. Adding further complexity to the characterization of these materials, the rheology of some fluids changes with time due to to their shear history and slow restructuring processes; this evolution is termed as thixotropy~\cite{thixotropy_BARNES19971, MEWISthixotropy,ThixotropyLarson}.  A detailed review of the behaviour of impacting drops of shear-thinning, yield-stress, and viscoelastic fluids can be found here~\cite{bertola2012bookchapter}.

Particulate suspensions are an important class of complex fluids, and their rheology can be conveniently tuned by changing the volume fraction, $\phi$, of particulate additives. At low $\phi$ these suspensions act as Newtonian fluids; shear thinning and thickening flow becomes more apparent as $\phi$ is increased [Fig.\ \ref{fig:rheo}b]. Shear thickening is considered a precursor to shear jamming~\cite{WyartCates2014}, where fluid flow is completely arrested, and solid-like behaviour is observed.  Close to shear jamming, the power-law description  [Eqn.\ \ref{powerlaw}] is no longer adequate, as the shear stress diverges.


Even in the low-$\phi$, Newtonian regime, the existence of particulate additives increases the overall measured viscosity, termed the \emph{effective viscosity} of the bulk suspension. This is a useful first approximation to compare complex fluid behaviour to its viscous Newtonian counterparts. In the low-$\phi$ limit, the effective viscosity of a suspension can be computed using the Einstein relation:
\begin{equation} \label{Einstein}
    \eta_\textit{\scriptsize{\mbox{eff}}}=\eta(1+2.5\phi),
\end{equation}
\noindent where $\eta$ is the viscosity of the surrounding fluid~\cite{Wagner2012colloidal}. This approximation holds well only in the dilute limit, and does not account for the suspension viscosity diverging at finite $\phi$ due to random close packing or `jamming' ($\phi_m=0.64$ for hard spheres). The Krieger-Dougherty equation~\cite{krieger1959mechanism} reflects this behaviour and accounts for the jamming  volume fraction; it is commonly used to predict effective viscosity over a much broader range of $\phi$:

\begin{equation} \label{KD}
    \eta_\textit{\scriptsize{\mbox{eff}}}=\eta(1-\frac{\phi}{\phi_m})^{[\eta^*]\phi_m},
\end{equation}

\noindent where $[\eta^*]$ is the `intrinsic viscosity' which is set by particle shape; $[\eta^*]=$2.5 for spherical particles. This relation captures two important behaviours: it reduces to Equation~\ref{Einstein} in the low-$\phi$ limit, and the viscosity it predicts diverges as $\phi\to \phi_m$, where the suspension jams and behaves as a solid material. By computing the effective viscosity in this manner, one can extend the definition of Newtonian dimensionless flow parameters to describe non-Newtonian flows, at least within some parameter regimes.

Below, some key dimensionless  parameters relevant to the work discussed here are defined, in forms they take for drop impact systems. Some of these parameters are defined for Newtonian fluids, and can be adapted for complex fluids. Some others pertain specifically to non-Newtonian systems. For the dimensionless numbers defined here:

$\rho$ is the density of the fluid,

$\eta$ is the dynamic viscosity of the fluid,

$d_0$ is the drop diameter,

$u_0$ is the impact velocity,

$\sigma$ is the fluid surface tension. 

\begin{itemize}

\item \textbf{Weber number ($We$)} is the ratio of inertial and surface stresses, $We=\frac{\rho u_0^2 d_0}{\sigma}$. Large $We$ signifies that surface stresses are negligible compared to inertia, while surface stresses dominate at small $We$.

For particulate suspensions, if the particles are large enough so that particle inertia dominates over the bulk fluid behaviour, the particle-based Weber number is useful: $We_p=\frac{\rho_p u_0^2d_p}{\sigma}$, where the particle diameter, $d_p$ is the relevant length scale.

\item \textbf{Reynolds number ($Re$)} is the ratio of inertial and viscous stresses. $Re=\frac{\rho u_0d_0}{\eta}$. At small values of $Re$ viscosity dominates, while large $Re$ implies viscous stresses are negligible compared to inertia. 

In case of non-Newtonian fluids, the \textbf{effective Reynolds number} is defined as $Re_\textit{\scriptsize{\mbox{eff}}}=\frac{\rho u_0d_0}{\eta_\textit{\scriptsize{\mbox{eff}}}}$, $\eta_\textit{\scriptsize{\mbox{eff}}}$ being the effective fluid viscosity. 

\item \textbf {Stokes number ($St$)} compares the viscous and inertial forces on a spherical particle of density $\rho_p$ suspended in a fluid, $St=\frac{\rho_p u_0d_p}{\eta}$. The stokes number is equal to the Reynolds number experienced by a single  spherical particle in a fluid. 

\item The \textbf{Ohnesorge number ($Oh$)} compares the effect of viscous stresses with the combined effect of surface stresses and inertia, $Oh=\frac{\sqrt{We}}{Re}$. The Ohnesorge number is suitable for systems where inertial, viscous, and surface stresses may all be relevant  --- a common scenario for drop impact at a few m/s.

\item The \textbf{Capillary number ($Ca$)} is the ratio of viscous stresses to surface  tension, $Ca=\frac{\eta u_0}{\sigma}$. $Ca$ is large for viscosity-dominated conditions, and small when surface tension dominates.

\item The \textbf{P\'eclet number}, $Pe=\frac{3\pi \eta \dot\gamma d_0^3}{4k_B T}$, where $\dot\gamma$ is the shear rate, compares the rate of advection by the flow to the rate of diffusion by Brownian motion in a suspension. For high values of $Pe$, the flow dominates over thermal motion, this is the high shear regime.

\item The \textbf{Elastic Mach number} ($M_e$) compares the fluid velocity to the elastic velocity of the fluid, $M_e=\frac{u_0}{\sqrt{G/\rho}}$,$G$ being the elastic modulus of the fluid. $M_e$ is especially relevant in viscoelastic fluid impact. 

\item \textbf{Weissenberg number} ($Wi$) compares the elastic forces in the system to the viscous forces.  Its definition depends somewhat on the system details.  For example, in steady shear flow, it is given by the ratio of the first normal stress difference to the shear stress: $Wi=\frac{\tau_{xx} - \tau_{yy}}{\tau_{xy}} = 2\lambda\dot\gamma$, where $\lambda$ represents the stress relaxation time of the fluid.

\item \textbf{Froude number} ($Fr$): compares fluid inertia to gravitational effects, $Fr=\frac{u_0}{\sqrt{gd_0}}$. For typical drop impact experiments, the Froude number is large; therefore we can ignore the effects of gravity in drop impact systems. 

\end{itemize}

\section {Key advances in Newtonian drop impact}
\label{Newtonian}

Despite the complexities of the drop impact process, a detailed understanding of impacting Newtonian drops has been built. Despite key differences between the flow properties of Newtonian and complex fluids, they share system details and experimental methods. Moreover, in certain regimes, the physics of impacting complex fluids can be understood using Newtonian models. Therefore, Newtonian studies provide the foundation for their complex fluid counterparts. In this section, we summarize the most pertinent results of Newtonian drop impact. \emph{We emphasize that this section is not intended to be a comprehensive review of Newtonian drop impact.} It is far too brief to cover more than a summary of this large body of work. We solely aim to highlight aspects of Newtonian impact that are most relevant to the impact dynamics of complex fluids. We focus on impact on a dry and stiff surface; a pre-wet~\cite{josserand2003droplet, Blackwell2015,sen_morales_ewoldt_2020}, compliant~\cite{howland2016s,pepper2008}heated~\cite{bertola2012bookchapter,BERTOLA2014259}, or otherwise complex~\cite{ZhaoPRL2017,GILET_BUSH_2009} impact substrate adds further complexity and can highly modify drop impact behaviour.  For a more complete discussion of Newtonian drop impact, we refer the interested reader to the following reviews on the subject~\cite{YarinReview, JosserandReview,KarimReview10.1063/5.0130043}.\par

After impacting a solid surface, a fluid drop radially expands on the timescale of  milliseconds. During expansion, fluid inertia is converted into surface energy, while being opposed by viscous stresses, fluid-substrate interactions, and ambient gas effects. In some impact regimes, the rim of the spreading drop becomes unstable, leading to ejection of secondary droplets, termed   splashing. After the spreading phase, the fluid may maintain its radially extended shape, or it may partially recede.  On hydrophobic surfaces, drops impacting at large impact velocities may fully retract and may even detach from the surface. A significant body of work has been done to understand spreading~\cite{laan2014maximum, lee_laan_bonn_2016}, receding~\cite{bartolo2005retraction}, bouncing~\cite{Quere, Bird2013contact},  as well as the transition from spreading to splashing~\cite{Riboux2014,YarinReview,JosserandReview}. Below, we summarize key findings from those Newtonian drop impact studies that most directly connect to the existing body of work for impacting complex fluid drops. \par 

Immediately after impact, a fluid drop undergoes inertial spreading. The maximum spreading diameter after impact is governed by the balance between drop inertia, surface tension, and viscous dissipation. Energy conservation arguments in the low-viscosity, high-inertia regime (large $Re$, $We$) predict the maximum drop spread to scale as ${We}^\frac{1}{2}$. However, in the high-viscosity, low-inertia regime (small $Re$), the maximum spread has been observed to follow the scaling of ${Re}^\frac{1}{5}$~\cite{JosserandReview}. It has thus been proposed that a broad crossover regime must exist between these two extremes. Laan et al.~\cite{laan2014maximum} used the first order Pad\'{e} approximation to propose the scaling:
\begin{equation}
    \beta_{max}= {Re}^\frac{1}{5}\left(\frac{P^\frac{1}{2}}{A+P^\frac{1}{2}}\right),
\end{equation}
\noindent where $\beta_{max}$ is the normalized drop spread, $d_{max}/d_0$, and $P=We{(Re)}^{-\frac{2}{5}}$ as suggested by energy conservation~\cite{Eggers2010,lagubeau2012spreading}. This scaling agrees well with experimental data for drops of glycerol-water mixtures of varying viscosities onto highly wettable substrates. A more recent scaling modifies this result to account for a range of surface wettabilitty~\cite{lee_laan_bonn_2016}, and finds agreement with experimental tests on a variety of surfaces.  The simple metric of maximum spread diameter can encapsulate much of the physics at play in drop impact, and captures the behaviour of a wide variety of fluids and substrates.  The recent  review by Josserand and Thoroddsen~\cite{JosserandReview} provides an in-depth discussions of these models and their comparisons to experimental data.  
  
After the spreading phase, the impacted drop may recede, rather than remaining at its maximum impact diameter. Bartolo et al.~\cite{bartolo2005retraction} observed that when impacted on hydrophobic surfaces at high $We$, the drop receding velocity was surprisingly independent of impact velocity. This suggests that while inertia governs the spreading phase, it has a negligible effect on the dynamics of the receding phase. The authors additionally observed that, consistent with simple hydrodynamic arguments, the receding rate depends on viscosity via the Ohnesorge number $Oh$. Subsequent numerical and experimental work has accounted for substrate wettability via the retracting dynamic contact angle~\cite{DuMin2021NumericalRetraction, wang2020retraction}. While substrate interactions only have a limited influence on droplet spreading, they can markedly alter retraction dynamics due to the much lower velocities (and thus inertial effects) in this process.  Experiments on small targets are effective for decoupling substrate interaction from fluid properties, and this technique has been used to explore spreading and splashing dynamics by many groups. Small target experiments have studied the spreading~\cite{arora2018smalltargets}, thickness evolution~\cite{vernay_2015_smalltargets}, and disintegration~\cite{smalltargets_newtonian} of expanding fluid sheets.  \par

In addition to receding, liquid drops impacting on hydrophobic surfaces may retract and then rebound, completely leaving the impact surface. For high $Re$ and $We$ impacts, drops bounce almost elastically, and have been successfully modeled as a simple spring-mass system~\cite{Quere}. Recently, this model was extended to additionally account for drop viscosity  using a  damped spring-mass equation~\cite{jha2020viscous}; this model also predicted an increase in the contact time with the substrate for more viscous fluids. Controlling bouncing and contact time is crucial for the development of water-repellent surfaces, and the use of micropatterns on the substrates has been proposed for such applications~\cite{Bird2013contact}. For a more detailed discussion of drop impact on hydrophobic surfaces, we refer the reader to this review~\cite{khojasteh2016droplet}.\par 

At the end of the spreading phase, a drop may splash. This is often an undesirable outcome, for example limiting the ability to smoothly and evenly coat a surface, and much work has gone into exploring how to suppress splashing.  The transition to splashing was originally proposed to be governed solely by fluid impact properties, and parameterized by the dimensionless number $K=We({Re})^{\frac{1}{2}}$~\cite{stow1981experimental,MUNDO1995151}. However, more recent studies have demonstrated that the quantitative value of the splashing threshold additionally depends both on the details of the substrate~\cite{RANGE1998} and surprisingly, the surrounding gas~\cite{Xu2005}. An impacting drop rapidly spreads over the substrate at m/s velocities, and thus understanding the details of how the liquid-solid contact point moves at very high speeds is crucial to studying this process.  A common approach in recent work is to characterise the motion of the contact line using the dynamic contact angle~\cite{latka2018drop, Quetzeri2019}; this fluid-solid contact is known to exhibit instabilities in other high-velocity coating flows~\cite{blake1979maximum}.  The dynamic contact angle during drop impact is set by fluid-substrate interactions, and differs during the spreading and receding phases~\cite{snoeijer2013}. 
It remains a challenge to measure this microscopic quantity experimentally~\cite{blake2023possible}, and a distinction is often made between an apparent or macroscopic contact angle and the microscopic dynamics of fluid-solid contact. Quetzeri-Santiago et al.~\cite{Quetzeri2019} proposed that the advancing dynamic contact angle (as opposed to the static contact angle) is the relevant parameter for predicting the splashing threshold, potentially extending our understanding of splashing from wetting to hydrophobic surfaces. 

The relevance of the dynamic contact angle to splashing processes has been
linked to the surprisingly strong influence of ambient pressure on splashing: Xu et al.~\cite{Xu2005} observed that lowering the ambient pressure could completely suppress splashing in ethanol drops.  This observation has been generalized to include a wide variety of liquids, and there is a growing body of work focused on understanding this counter-intuitive result~\cite{Riboux2014, Riboux2017, deGoede2018,latka2018drop,jian2018,Quetzeri2019}. At early stages of contact, an air layer has been observed under the spreading fluid~\cite{Kolinski2019, Thoroddsen2003, Driscoll2011, Liu2015, vanderVeen2012}, and this air layer has been proposed to be linked to the pressure-dependence of the splashing threshold, though this is a point of debate in the community, especially for more viscous fluids at higher impact velocities. Recent work has focused on incorporating the effects of gas viscosity, density, and mean free path on splashing~\cite{Riboux2014, Riboux2017, jian2018}. The work so far has neglected gas compressibility; incorporating this effect in future studies would be valuable, albeit challenging.  \par


Although most drop impact studies have focused on normal impact and hard substrates, real-life impacts often happen at an angle, or on compliant surfaces. The best example of this is the spraying of pesticides onto leaves, a process where both of these modifications come into play. Studies of splashing onto oblique and translating surfaces~\cite{bird2009, Hao2019, Garcia2020, laan2014maximum, bird2009} have observed an asymmetric splash; this bifurcated splash is a reflection that the spreading drop experiences a varying normal component to its impact velocity when it hits a translating or tilted surface. In this case, standard models for maximum spreading diameter are still applicable when the normal component of impact velocity and short axis of spreading~\cite{laan2014maximum} are used. Drop impact work on compliant substrates~\cite{pepper2008, SoftSplash} has reported suppressed splashing (as compared to a rigid substrate), and linked this suppression to the increased energy dissipation on a softer impact substrate~\cite{SoftSplash}. Gilet and Bush~\cite{GILET_BUSH_2009} have given a detailed experimental and theoretical treatment of Newtonian drops impacting a fluid `trampoline' of soap films, leading to rich periodic and chaotic behaviour. Newtonian drop impact on granular beads has also been studied ~\cite{ZhaoPRL2017}. While a modified Weber number relates the transition between spreading and splashing regimes is similar to smooth substrates, nuances of air layer dynamics and shear banding require further exploration.  Studies on these lines would be relevant to agriculturally pertinent questions such as  the permeation of raindrops into soil.  We direct readers to this review~\cite{KarimReview10.1063/5.0130043} for a detailed discussion of Newtonian drop impact on substrates of varying wettability, roughness, and hardness.\par 

Although the properties of complex fluids are significantly different from Newtonian fluids, many parallels can be drawn due to identical system details such as the free-surface geometry and substrate properties. Broadly defined behaviours such as spreading, receding, splashing, and bouncing have counterparts for impacting complex fluid drops, although the quantitative trends and governing parameters can be quite different. In the following sections, we aim to connect disparate observations of impacting complex fluid drops to the existing foundation of Newtonian studies. Throughout, we highlight phenomena that need further exploration in order to build a more unified picture of the physics of non-Newtonian drop impact.

\section{An outline of experimental and computational methods}
\label{methods}

\subsection{Experimental methods}

The essential elements of a drop impact experimental setup are: a substrate, a source that produces drops, and a light source and camera to image the impact. The simplest setups involve a horizontal substrate and a needle connected to a reservoir to produce drops. The needle is positioned at a fixed height above the substrate, and when the drop falls, it is allowed to accelerate under gravity; the post-impact dynamics are recorded using the light source and the camera. This basic setup can be modified in a variety of ways to explore additional impact outcomes. For example, the  substrate can be positioned at various angles or made to move to adjust the horizontal component of the impact velocity. The needle can be replaced by more industrially relevant drop sources such as a spray gun. The fundamental mechanics of the experimental setup remain unchanged for both Newtonian and complex fluid drop impact~\cite{YarinReview, JosserandReview}. However, as complex fluids can dramatically change their flow properties with the time-varying impact stresses, it is important to take this into account when interpreting impact outcomes.

The bulk of this review discusses the most straightforward impact scenario --- impact dynamics on smooth, dry substrates. It is important to note that the hydrophilicity of the substrate may drastically change the impact dynamics. To address this challenge, most experimental work focuses on a limiting case, employing either largely hydrophilic or largely hydrophobic substrates.  The most commonly used hydrophilic substrate is a smooth glass slide. Impurities such as dust, fingerprints, and organic contaminants on the substrate can greatly alter its hydrophilicity, hence a thoroughly clean substrate is needed for controlled experiments. Glass slides are commonly cleaned either by plasma cleaning or by washing them with a concentrated base solution and water.  These cleaning procedures ensure the slides are highly hydrophilic, e.g.\ water drops will have a contact angle of a few degrees. In the other limit hydrophobic substrates can be realized using a variety of techniques.  These surfaces can  be made from a bulk hydrophobic material such as  PTFE or PDMS, be fabricated from glass or a silicon wafer coated with a layer of hydrophobic material, or micropatterned with a texture which creates  hydrophobicity on the macroscale. Regardless of substrate kind, surface treatment, and the type of fluid under study (Newtonian or otherwise), it is crucial that the substrate is characterised using reference liquids such as water and ethanol so that wettability is a well-controlled parameter. 

Working with non-Newtonian fluids introduces additional challenges to preforming and interpreting drop impact experiments.  To create Newtonian droplets, a needle and syringe pump is often employed to form drops in a simple but highly controlled fashion. However, this method cannot always be used for complex fluids.  In cases where the complex fluid under study is high in viscosity, susceptible to sedimentation, drying and clogging, or retains a memory of shear, use of a syringe pump is not ideal. Additionally, for highly specialized complex fluids synthesized in the lab, available sample volume becomes  another constraint. Addressing both of these challenges, an alternative drop creation technique is to manually form a drop using a micropipette~\cite{Shah_2021_impact}. Although experimentally tedious, this method is efficient in terms of amount of fluid used , allows for thorough mixing of the bulk fluid between drops to minimize sedimentation effects, and mitigates the risk of drying and clogging. Another challenge in creating non-Newtonian drops is that the shear experienced during drop formation may affect the shape and rheological properties of the drop formed;  this is especially pronounced in highly concentrated particulate drops. Although impossible to completely eliminate, forming the drop quasi-statically at a known rate is helpful to control this unwanted shear history. The timescale of sample preparation may affect experimental data due to changes such as drying, phase-separation, and thixotropy over time. When it is not possible to prepare samples right before experimentation, care should be taken with sample storage, and re-mixing may be necessary right before experimentation. 

Interpreting the behaviour of complex fluids at impact introduces additional challenges.  As complex fluid viscosity is a non-constant function of shear, the flow of complex fluids under impact is often understood using  flow curves from steady-state rheometry, and the appropriate viscosity is to be used is chosen by estimating the shear experienced by the drop during impact. Alternatively, the fluid viscosity may be theoretically estimated by using approximations such as the Einstein equation for dilute suspensions, or viscosity may be measured using standard viscometers. For viscoeleastic and yield-stress fluids, oscillatory rheology may be used to characterise the fluid~\cite{Wagner2012colloidal,shaw2012introduction,macosko1994rheology}.  In each case, great care should be taken to ensure an appropriate impact flow behaviour is used for any modeling.

Dense particulate suspensions are particularly  challenging to work with experimentally, largely because their rheology and impact dynamics are extremely sensitive to the volume fraction of particles, which is a challenging parameter to control with high precision. Moreover, drying effects (fluid loss due to evaporation on the experimental time scale) are significant enough to change the volume fraction, and subsequently the impact dynamics of dense suspensions. Drying effects may compete with the longer-time spreading dynamics of complex fluids. These challenges necessitate precise control and monitoring of humidity during impact experiments with dense suspensions. The entire path of the drop from its formation to impact may be enclosed in a chamber where constant humidity is maintained either by inserting a small reservoir of water, or by continuously circulating humid air. The humidity level should additionally not be too high; the air inside the chamber becoming saturated with water vapor may inadvertently create a substrate coated a thin layer of water, and change the experimental parameters. Therefore, it is best practice to monitor the humidity in real time. 

The most common avenue of data collection for impacting drops is high-speed imaging. Details of the imaging setup are largely independent of the fluid under study. The opacity of dense particulate or polymeric fluids may introduce challenges to more advanced methods such as particle tracking and velocimetry to understand the dynamics inside the drop. Index-matching might mitigate some of these issues, but they may introduce constraints on the kind of suspending liquid. Sophisticated light sources such as laser sheets or highly monochromatic light might be needed to image inside the drop or studying the dynamics underneath the drop. In addition to imaging data, the force on the substrate may be measured. This provides useful information, but places limits on the type of the substrate used.

\subsection{Computational methods}

Drop impact studies have been historically experimentally-driven due to the challenges in modelling this process.  The key reason that drop impact dynamics are challenging to reproduce with numerical methods is that this is a multiphase flow problem, e.g., one must capture the dynamics of a liquid moving through a gas and interacting with a solid (or another liquid).  Moreover,  a large range of both time and length scales are involved. Modeling of non-Newtonian fluids only amplifies this challenge, as the rapid and large change in velocity from the moment of impact to late spreading results in marked changes in flow behaviour.  While it is important to understand the limits of modelling this complex problem, multiple tools have been adapted to explore this problem.  
Computational fluid dynamics is a large field, and it is not within the scope of this review to describe these techniques in depth. We instead refer the interested reader to the following texts~\cite{ferziger02:CMFD, hughes2012finite, anderson2020computational, kruger2017lattice}.  Below, we aim only to highlight the most common techniques and discuss in general their advantages and disadvantages.  

A commonly employed technique, especially in the engineering community, is using solvers based on the finite element method (FEM)~\cite{glowinski1992finite, kieckhefen2020possibilities}.  There are numerous software packages, for example COMSOL, ANSYS, and OpenFOAM, which exist that employ these methods, and can aid in dynamic meshing, addressing numerical instabilities,  and other challenges that arise when modelling moving fluids.  These methods are well-validated on general problems, can have a low computational cost, and have a reasonably low barrier to entry.  However, in all meshed techniques, care must be taken when modelling dynamic processes such as drop impact, as processes occurring at small scales (development of the lamella for example) must be resolved by the mesh, and otherwise can have a large influence on the overall process, and control splash outcomes. FEM typically employees an irregular, body conforming mesh (constructed from triangles and/or quadrilaterals), which is robust but unwieldy; every time the drop changes shape, a complex and computationally expensive remeshing problem must be solved across the whole domain.  Alternative techniques typically use regular gridding, and only modify it near the fluid interface, which greatly can help alleviate the computational expense of remeshing.  In all cases, extensive validation should be done to ensure these methods correctly capture the physical processes during drop impact.

In addition to the complications introduced by using a mesh, explicitly solving physics across the entire domain of the problem is computationally expensive.  Thus, a variety of alternatives to the FEM approach have been developed for multiphase flow problems such as droplet impact which instead focus on evolving dynamics by tracking the position of the fluid interfaces.  These alternatives typically fall into two categories, interface tracking methods, where the fluid interfaces are explicitly tracked and advected, and interface capture methods where instead the interface is implicitly followed, and is computed as a contour of a particular scalar function.  

The main interface tracking method which has been used to study drop impact is the volume of fluid (VOF) method~\cite{gueyffier1999volume}; examples of VOF packages commonly employed for drop impact include Basilisk and OpenFOAM.  VOF exploits numerical techniques that track the shape and position of the interface rather than explicitly solving for dynamics at every point in the fluid. Therefore, VOF can overcome difficulties due to stiffness in the underlying equations or numerical instabilities, and more importantly largely employs a regular (not body-conforming) mesh, which leads to a large improvement in computational efficiency.  However, VOF is still a meshed technique, and can suffer from inaccuracies in estimating the surface tension force, slow numerical convergence, and correct capture of sharp interfaces which can lead to unphysical results.  As with FEM, care should be taken to validate flow predictions.  An alternative approach is to use an interface capture method; the chief methods employed for drop impact are phase-field modeling~\cite{Kim_2012} and Level Set methods~\cite{sussman1994level, osher2001level, osher1988fronts}. Phase-field models represent the interface as a contour of a scalar function (the phase field); the sharp discontinuity in physical properties at a fluid interface is replaced by transition region.  This technique can be quite computationally efficient as the interface is not explicitly tracked. However, this diffuse interface can become a disadvantage and lead to inaccuracies in high-curvature regions, which can lead to quite stringent time-step requirements.  As implied by their name, Level Set methods represent the interface as the level set of a shape and then follow its evolution in time.  In contrast to phase-field methods, Level Set methods are particularly adept at capturing sharp interfaces and topological changes, but it can be quite challenging to construct the appropriate velocity function for advecting the level set. 

Two other mesh-free approaches have been used to numerically study multiphase flow problems such as drop impact, the Lattice Boltzmann Method (LBM) and Smoothed Particle Hydrodynamics (SPH).  LBM is based on a gas kinetic approach~\cite{gunstensen1991lattice,aidun2010lattice,chen1998lattice}.  It does not explicitly solve the Navier-Stokes equations, but instead using a microscopic, probabilistic description of fluid particle interactions (the Boltzmann equation); the evolution of this probability density function is then solved on a discretized lattice.  LBM are computationally efficient, easy to parallelize, and especially amenable to complex solid boundaries; they have had great success correctly capturing many kinds of multiphase flow dynamics.  However, LBM is only explicitly shown to approach the physics given by the incompressible Navier-Stokes equations in asymptotic limits; this places restrictions on which flows can be captured by the method, and extensive validation is required.  SPH~\cite{pozorski2023smoothed,wang2016overview} represents fluid elements as particles, with the fluid constitutive relation providing the particle-particle interaction, and is implemented in some common solvers such as ANSYS; flow dynamics are then obtained by interpolating over all of the particles.  SPH has the tremendous advantage of being a mesh-free method, and thus is ideally suited for solving problems with complex boundary dynamics such as splashing.  However, due to the large number of particles needed it can be computationally expensive, and care must be taken during interpolation.  

Many techniques exist for numerically studying drop impact, and all have inherent advantages and disadvantages.   Due to imaging challenges, high-resolution flow field information is extremely difficult to obtain experimentally in this system and numerics provide an important tool to understand impact dynamics.  While drop impact remains a challenging problem from a computational prospective,  great progress has been made in the past decade.    When used with care, computational tools are invaluable for enhancing our understanding of this complex process.

\section{Polymeric fluids}\label{polymeric}

Polymer additives are used in a large variety of industrial applications, both to achieve desired fluid properties, for example increased viscosity, and to control the interaction of fluids with solid substrates. In section \ref{PolyExp}, we outline the experimental results in polymeric drop impact, highlighting connections between material rheology and drop impact outcomes. The two main parameters varied in experiments on polymeric fluids are the polymer concentration and the molecular weight (or chain length) of the polymer. Changing either of these can alter multiple rheological properties of the fluid, for example the yield-stress, the shear-thinning coefficient, and the zero-shear viscosity. Numerical studies have proven beneficial to study the effect of each of these properties separately. We describe in Section \ref{PolyNum} numerical  and modelling work in this field, and then conclude the section by highlighting open questions and potential future directions.  \par

\subsection{Experimental results}\label{PolyExp}

\begin{figure*}[t]
\centering
\includegraphics[height=12cm]{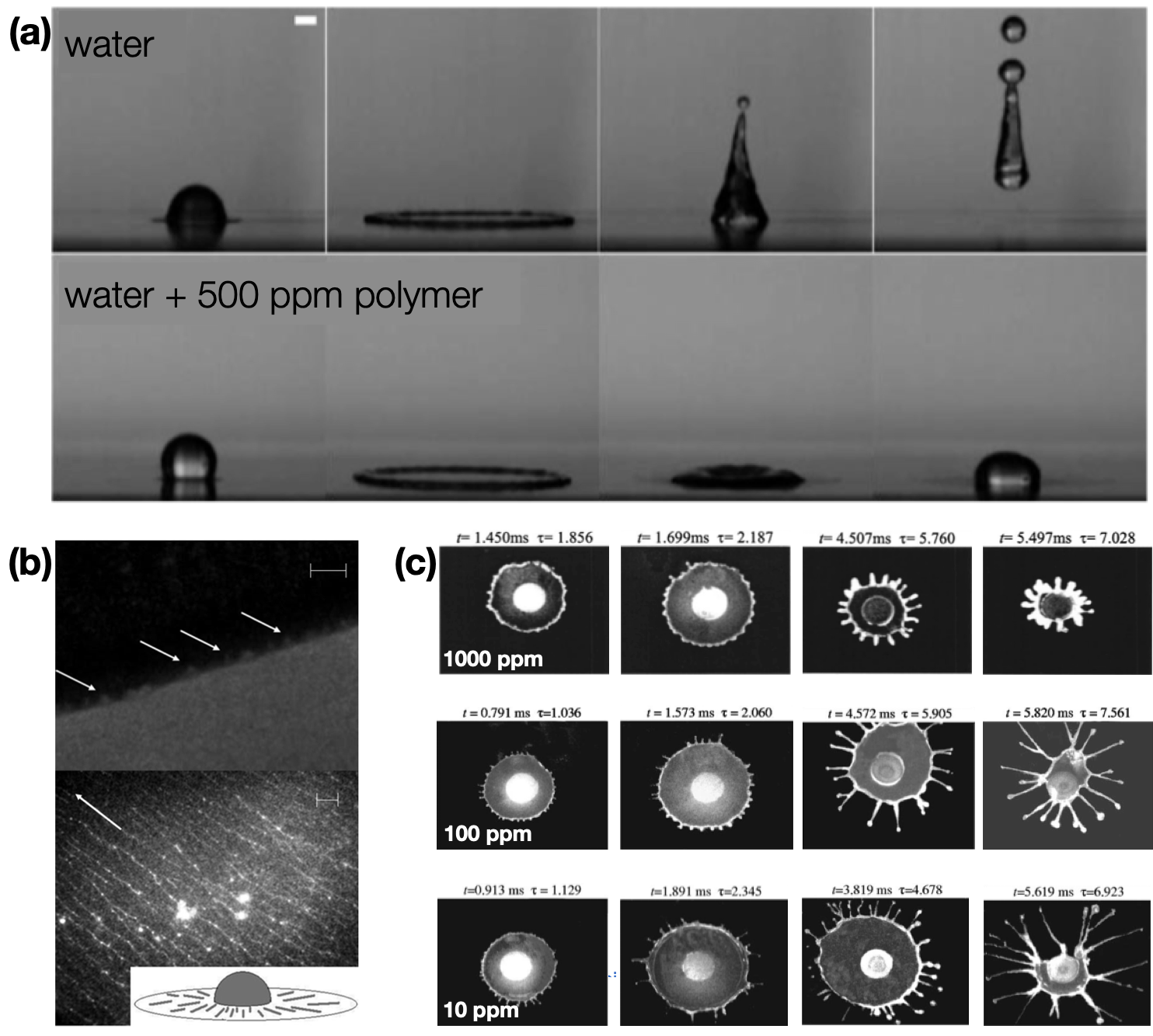}
    \caption {\textbf{Drop impact processes can be substantially altered by even small concentrations of polymer additives. }(a) A small concentration of polymer additive can completely suppress  droplet rebound.  Adapted from~\cite{Bartolo2007}.  (b) Demonstration of polymer (DNA) extension via the moving contact line during the receding phase, as illustrated by the sketch.  Upper panel: DNA protruding from the contact line during receding (scale bar 20 $\mu m$ ).  Lower panel: The DNA deposited on the substrate by the receding drop is highly stretched, and aligned in the direction of flow. Adapted from~\cite{SmithBertola2010}.  (c) Timelapse of a polymeric drop after impact on a small target.  The polymer additives act to suppress both edge instabilities and splashing, and this effect becomes more pronounced as polymer concentration is increased. Adapted from~\cite{Rozhkov2003}
}
    \label{fig:polymer}
\end{figure*}

Polymer-based materials, being widespread in natural systems, are common fluid additives.  For industries such as coating, spraying, and pesticide dispersal, suppressing bounce and splash is crucial for the efficient use of products, and polymeric additives can greatly aid in this effort. Impact on hydrophobic substrates is of special interest in agrochemical applications, as natural substrates like leaves often have microstructure that results in varying degrees of hydrophobicity. Polymer additives are a remarkably efficient way to modify impact processes [Fig. \ref{fig:polymer}(a)]; Bergeron et al.~\cite{Bergeron2000} observed that a very small amount ($\sim$ 100 ppm) of polymer additives completely suppressed the bounce of liquid droplets impacting on a variety of hydrophobic surfaces. Bounce suppression is more pronounced for higher molecular weight polymers~\cite{CrooksCooperWhite2001}, and such a suppression has since been reported by many other studies~\cite{Bartolo2007,SmithBertola2010,Zang2013,CrooksBoger2000}. On the contrary, according to Huh et al. ~\cite{huh2015role}, this suppression is only observed when the polymer concentration exceeds 0.03\%. In addition to concentration, the polymer chain length, conformation (linear vs. branched) and chemical functionality may have an effect on bounce suppression, and these parameters need to be investigated in more detail in future experiments.

To suppress droplet rebound, additional energy must be dissipated during the droplet impact process when polymer additives are present.  To understand the mechanistic origin of this dissipation, it is necessary to understand how polymer additives modify fluid properties.  Due to their elongated nature, polymeric materials can often have quite different effects on the different components of stress.  Rheological data suggests that small concentrations of polymer additives largely do not affect fluid shear viscosity. However, under extensional shear polymer chains  elongate and deform, giving rise to an increased elongational viscosity.  As the fluid flow during both drop spreading and retraction is largely elongational, bounce suppression was initially attributed to this increase in this elongational viscosity~\cite{Bergeron2000,CrooksBoger2000}. Later studies have disputed this hypothesis, arguing that in addition to bounce, both spreading and receding should be affected by a higher elongational viscosity~\cite{Bartolo2007}, and this contrasts with experimental results.  In particular, it is known that a low polymer concentrations, the spreading phase is unmodified, while the retraction phase occurs at a suppressed velocity. However, in highly concentrated solutions, velocity suppression in both the spreading and the receding phases has been observed~\cite{GelledFuel}. While in past work normal stresses have been proposed to play a role in bounce suppression,  a more thorough analysis has shown this to be an unlikely factor to explain this phenomenon~\cite{BERTOLA2013polymerimpact}.  In addition to modifying bulk flow properties, polymer additives modify liquid-surface interactions at the drop contact line, and recent work suggests that, as with Newtonian fluids, contact line dynamics are crucial to understanding polymeric drop impact~\cite{BERTOLA2013polymerimpact}.  Particle Image Velocimetry data by Smith and Bertola~\cite{SmithBertola2010} indicates that the contact line velocity is lower in impacted polymeric drops. This points to increased dissipation at the contact line, rather than bulk flow properties, playing a key role in bounce suppression. Smith and Bertola additionally found that this slower velocity was correlated with the stretching of individual polymer chains [Fig.\ \ref{fig:polymer}(b)].  Further experiments focusing on altering contact line dissipation through substrate modifications would shed more light on this phenomenon, and potentially establish whether this mechanism is universal, or is specific to long and highly flexible polymers.

While the receding phase is greatly modified by the addition of polymers, the spreading phase is nearly unchanged~\cite{huh2015role}, and Newtonian models can quantitatively describe the spreading of shear-thinning polymeric drops.  An and Lee~\cite{AnLee2012} studied this regime using xanthan gum solutions, and compared their behaviour to Newtonian drops. They observed that the maximum spreading diameter of the shear thinning liquid was qualitatively similar to that of Newtonian fluids, provided one used a fluid viscosity defined by an average of the infinite shear viscosity and the zero-shear viscosity. Thus, by appropriately quantifying the average shear viscosity for the shear-thinning liquid, one can capture its spreading behaviour using existing Newtonian models~\cite{SchellerBousfieldNewtonian, laan2014maximum}.

Similarly to Newtonian drop impact~\cite{arora2018smalltargets, vernay_2015_smalltargets, smalltargets_newtonian}, the interaction of the impacting drop with the substrate can be decoupled from fluid behaviour by impacting the droplet onto a small target, so that it only interacts with the substrate for a brief time at the beginning of the spreading process. Given the implication of contact-line dissipation in other impact outcomes, such as bounce suppression, this has been a particularly active area of study.  Rozhkov et al.~\cite{Rozhkov2003} observed that the spreading of polymer solutions on small targets to be similar to that of water. This supports the assessment that altered retraction dynamics are due to polymer interactions with the substrate~\cite{SmithBertola2010, BERTOLA2013polymerimpact}.  It has been additionally observed that polymeric fluids exhibit suppressed splashing, and a reduction in edge instabilities and more stable ejected films [Fig.~\ref{fig:polymer}(c)].  While these results are consistent with the stabilizing effect of polymers on splash and bounce observed in other studies~\cite{Bergeron2000,Bartolo2007,SmithBertola2010,Zang2013,CrooksBoger2000}, these observations are unlikely to be due to surface-mediated dissipation, highlighting the wide array of physics that drives the behaviour of a splashing drop.

In addition to modifying fluid shear and elongational viscosity, polymer additives can impart an elastic response, which can play an important role in  impact dynamics. Most surfaces encountered in industrial applications are not smooth, but instead are rough and/or porous; many studies of polymeric fluids have focused on how surface modifications are coupled to impact outcomes.  Lee et al.\ explored the impact of viscoelastic fluids (Xanthan gum solutions) onto mesh surfaces, and correlated the penetration of the fluid into the mesh with the rheological properties of the fluid~\cite{lee2023rational}. Luu and Forterre~\cite{luu_forterre_2009, LuuForterre2013} observed that drops of carbopol, a yield-stress fluid, `super-spread', that is, they spread to a much larger extent on rough substrates; the spread was found to be even larger on hydrophobic surfaces. Their data scaled well when represented in terms of the elastic Mach number, $M_e$, indicating that the fluid elasticity is the dominant mechanism for determining the impact outcomes  in this system.  
They also suggested that for rough surfaces and high Weber numbers, the impact timescale was too short for fluid to seep through the surface troughs, and thus the super-spreading on rough surfaces was due to aneffective reduction in the contact area and in turn, frictional dissipation. Guemas et al.~\cite{Guemas}, on the contrary, did not observe such super-spreading on surfaces of similar roughness. The only different parameter between these studies was drop diameter. Hence to understand this phenomenon better, more experimental data as a function of drop size and roughness is needed. Microscopic imaging and experiments with patterned 3-d printed substrates might also shed light on the dynamics of this process. \par

Even though a number of experiments have focused on impacting drops on dry surfaces, drops often impact on an already wet surface in industrial processes. Blackwell et al.~\cite{Blackwell2015} observed that splashing was suppressed for higher concentrations of carbopol when impacted on wet surfaces. The ratio of coating thickness on the substrate and drop diameter emerged as a key dimensionless factor to demarcate the splashing regime. Another relevant parameter was the ratio of inertial and dissipative stresses. 
In a further study~\cite{sen_morales_ewoldt_2020}, the effects of thixotropic aging in Laponite on drop impact were explored, and these results could be characterised using the same dimensionless groups. Although these dimensionless parameters were successful in interpreting the data in this case, modifications need to be made for systems where surface tension and elasticity effects are significant. In a deep pool of liquid, splashing can occur due to movement of the bulk liquid; this is often termed `crown splash' in Newtonian liquids~\cite{YarinReview}. Recent work explored the impact of Newtonian droplets into pools of shear thinning (xanthan gum) and viscoeleastic (PEO) liquids, and found that the elasticity imparted by polymer additives could alter or even suppress instabilities in the pool after impact~\cite{mohammad2019experimental} .  Pre-wet surfaces are nearly ubiquitous in industrial applications such as spray coating and pesticide dispersal, and many of the fluids employed for these uses have polymeric additives.  Exploring the full phase space of how liquid layers modify impact outcomes is critical to efficient use of these materials.  Given the complexity of these impacts, this work must be experimentally driven, with a focus on connecting fluid rheology to impact outcomes.  More experimental data spanning a larger range of both liquid layer thicknesses and fluid properties (e.g., sampling a wide range of constitutive relations) should be collected to obtain a complete picture of polymeric drop impact on wet surfaces.\par

In summary, the presence of polymer additives has a stabilizing effect on impacting drops, that manifests through suppression of bounce and splash~\cite{Bergeron2000,Bartolo2007,SmithBertola2010,Zang2013,CrooksBoger2000, CrooksCooperWhite2001}. Studies on wet surfaces have also shown that liquid layers on substrates enhance this stabilizing effect~\cite{Blackwell2015, sen_morales_ewoldt_2020}.  Additionally, the super-spreading phenomenon observed by Luu and Forterre could have fascinating applications for coatings and should be further explored; a large body of systematic experiments on substrates of varying roughness and with drops of different sizes is needed in order to understand the parameters governing this surprising behaviour. In the shear-thinning regime, the spreading of polymeric drops can be captured by Newtonian models, provided the effective fluid viscosity is defined as the average viscosity~\cite{AnLee2012}. Many mechanisms have been suggested for bounce suppression in low-concentration polymeric drops, but a strong consensus is yet to emerge. Experiments that incorporate direct measurements of shear, elongational, and normal stresses~\cite{ChengStressReview} during spreading would greatly enhance our understanding of polymeric fluid impact. These measurements are quite challenging in practice, and thus numerical work has been useful tool to understand these processes, as we discuss below. \par

\subsection{Numerical studies}\label{PolyNum}

In experiments involving polymeric drop impact, the natural choice for the independent parameter is the concentration of polymer additives. The fluid's yield stress increases with increasing polymer concentration, as does the effective viscosity and rate of shear thinning. This coupling therefore makes it challenging to experimentally disentangle the effects of these different fluid properties on impact behaviour. Thus, numerical studies are of great value for separately understanding the effects of yield stress and flow rheology. Computational work has also provided valuable insights on local variation in fluid behaviour, which is significantly more challenging to capture in experiments. \par

Kim and Baek~\cite{KIM201262} performed a numerical study of impacting yield-stress drops, where the fluid rheology was modeled using the Herschel-Bulkley constitutive equation~\cite{Herschel1926}. To study the effects of each modification to the fluid properties on the spreading and retraction phases, they separately varied the yield-stress, viscosity, shear-thinning rate, and surface tension. We emphasize that these parameters are not possible to vary independently in experiments, illustrating the strong role numerical simulations can play in studying impact processes.  The spreading phase was found to be dominated by drop inertia and effective viscosity, and largely unaffected by surface tension and yield stress. The receding phase, on the other hand, was inhibited by both yield stress and capillarity. Experimental reports that high polymer concentrations modify both the spreading and retraction phase of drop impact are  consistent with this data, as the effective viscosity and yield stress change simultaneously with polymer concentration. However, a quantitative agreement with the experimental measurement of the spreading and retraction rates was not found. The variation in the dynamic contact angles during the spreading and retracting phases was not accounted for in this study, likely leading to the quantitative deviation.  

Recently, Oishi et al.~\cite{oishi_thompson_martins_2019} investigated both normal and oblique impacts numerically. Thixotropic effects were accounted for by introducing a delay between applied stresses and the resulting structural changes in the fluid. The study considered the phase space of elastic Ohnesorge number $Oh_e$, and a parameter $Y$ consisting of the elastic Mach number and the thixotropic timescale. This phase space was clearly divided between the sticking, bouncing, and rolling (in case of oblique impacts). These numerical results qualitatively matched the experimental data of Luu and Forterre~\cite{luu_forterre_2009}. However, in this study, Oishi et al.\ ~\cite{oishi_thompson_martins_2019} treated the thixotropic timescale as a fitting parameter for their numerical data. Theoretical development of physical models exploring this timescale would lead us closer to a clearer understanding of thixotropy in complex fluids. 

Existing numerical work has thus shed light on potential mechanisms by which polymeric additives can modify the spreading and receding phases of droplet impact on dry substrates.  In many coating and spraying processes,  polymeric drops impact onto wet substrates; computational studies focused on this (albeit more challenging) regime would help explore a wider range of liquid layer thicknesses.  Numerical work in this direction would bridge experimental data with physical insights of impact behaviours such as sticking and bouncing, and their relation to the  many and complex interactions between an impacting  drop and the substrate. Fitting parameters in existing studies need more theoretical attention, so that they may be connected to physically meaningful parameters that can be used to guide application design. Numerical studies are well-positioned to explicitly test the suggested mechanisms of bounce suppression, and bring clarity to the physics involved. Finally, the link between non-uniform stresses in the drop and spreading/receding behaviours is likely best explored in a numerical context.

\section {Particulate suspensions} \label{particulate}

Traditionally, particulate suspensions are divided into two broad classes based on particle size: Brownian and non-Brownian suspensions. Brownian suspensions are made of particles that undergo Brownian motion in a water-like fluid at room temperature, i.e.\ thermal effects dominate over particle inertia. We note that in various communities, the terms `Brownian suspensions' and `colloidal suspensions' are often used interchangeably. Fluids containing even a small amount of nanoparticles are known to have drastically different spreading-splashing behaviour after impact, which cannot be fully captured by modifications to bulk properties such as effective viscosity and surface tension.~\cite{AKSOY2022nanofluid}. Non-Brownian suspensions (sometimes referred to as `granular suspensions'), on the other hand, consist of particles too large to undergo significant random motion due to thermal effects. The particle size that divides these categories is often given as $\sim10$ \textmu m. However, the transition between these two regimes is not well-defined, and suspension behaviour may depend on many other factors such as the properties of the suspending fluid, the relative density of the particles in the fluid, and the flow velocities in the system. Moreover, drop impact processes happen at high P\`eclet numbers, where one expects that the contribution of thermal diffusion to impact dynamics may be negligible. However, particle inertia can still play an important role in the dynamics of impact, when compared to bulk fluid properties. In fact, for large particle additives, particle inertia can dominate over the bulk fluid rheology, and the impact dynamics are best evaluated using particle-based parameters. Thus, for the purposes of this review, we chose to classify drop impact studies into two regimes: one in which impact dynamics are governed by the bulk fluid rheology, and one in which particle inertia governs impact outcomes. We emphasize that our discussion here is not classified strictly on the lines of particle size, but instead on the observed fluid behaviour in the context of all the system parameters. Section~\ref{fluid} discusses studies where impact behaviour can be connected to the bulk fluid rheology, whereas Section \ref{particle-based} discusses studies where the inertia of individual particles governs impact outcomes.

\subsection{Bulk rheology-dominated regime} \label{fluid}


A drop impacting on a solid substrate experiences large shear rates that vary spatially throughout the drop as well as over time. This spatiotemporal variation is especially relevant to the problem of impacting particulate suspensions, as their rheological properties can change dramatically with small changes to the shear rate, $\dot\gamma$, in addition to the particle volume fraction, $\phi$ (see Fig\ \ref{fig:rheo}b). Below, we highlight impact outcomes of particulate suspensions in the regime that is governed by bulk rheology, and draw connections to insights from rheology data. \par

Particulate suspensions have complex rheological properties, and can exhibit both shear thinning and shear thickening behaviours in certain ranges of $\phi$ and applied shear~\cite{Wagner2012colloidal,DennMortonSoftMatter}. In the dilute limit ($\phi\lesssim 0.1$), these suspensions behave quantitatively similar to a Newtonian fluid, in that their viscosity is constant with respect to shear rate; in this limit particulate additives only increase the effective suspension viscosity. At higher $\phi$, shear thinning is apparent as a decrease in the fluid viscosity as applied shear is increased. At high $\phi$ and shear rates these fluids are shear thickening, so that the fluid viscosity increases with shear. The same fluid can exhibit both shear-thinning and thickening behaviours depending on the applied stress; dense suspensions often exhibit shear thinning at low shear stresses and then begin to shear thicken as the shear stress is increased.  A complete microstructural understanding of the mechanisms underlying shear thinning and thickening remains elusive, though the community has made great progress. The emerging consensus is that shear thinning is coupled with shear rearrangement of particles, while shear thickening may be a result of a transition from hydrodynamic to frictional interactions between particles.  We note that this understanding is still a matter of debate within the suspensions community, and that many factor are at play in the thickening mechanism in particular, for example particle size, roughness, and interparticle interactions.  For a more in-depth discussion of the mechanism underlying these rheological behaviours, we refer the reader to these recent reviews: \cite{morris2020toward, ness2022physics, DennMortonSoftMatter}. 

Particulate suspensions generally exhibit shear thinning over a wide range of $\phi$. Although one would expect shear thinning to significantly modify post-impact spreading, the drop diameter $d$ during spreading has been shown to grow in a manner identical to the spreading of Newtonian drops, provided an appropriate viscosity value is chosen. Multiple studies over a range of particle sizes~\cite{Guemas,German_2009, Shah_2021_impact,jorgensen2020deformation,Ok,nicolas_2005,Boyer, Raux, Marston, Grishaev, GRISHAEVdist} have reported that the spreading of suspension drops can be effectively quantified using spreading models for Newtonian fluids~\cite{SchellerBousfieldNewtonian,laan2014maximum}. To model the data in this way, one must compute an effective viscosity for the suspension, as this is an input parameter for these models; quantifying the effective viscosity is non-trivial, as the spreading velocity, and thus the shear rate is continuously changing in time.

Theoretically, one could calculate the bulk effective viscosity of the suspension as an extension of the Einstein viscosity beyond the linear term~\cite{krieger1959mechanism}. In practice, however, this calculation is quite sensitive to the particle concentration, and the experimental value inferred from the rheological data is often used. All studies of particulate suspensions find that as the effective viscosity of the suspension grows with increasing $\phi$, the maximum spreading diameter after impact, $d_{max}$, decreases. Even for particles over a $100$ $ \mu m$ in size, the spreading of particulate suspension drops has been directly compared to the spreading of Newtonian drops with a similar viscosity~\cite{jorgensen2020deformation}, despite the decidedly non-continuum nature of such a suspension. Thus, experimental measurements have clearly established that the bulk effective viscosity is a useful control parameter to quantify drop spread for particulate suspensions, even well beyond the Brownian limit.\par

\begin{figure*}[h]
\centering
\includegraphics[height=3cm]{Figures/suspension_figure_2.png}
    \caption {{Impact behaviour of suspensions.} (a) Suspension drops at low to moderate $\phi$ exhibit spreading behaviour that is quite similar to Newtonian liquids, but  the retraction phase is strongly influenced by particle additives of different sizes.  Adapted from \cite{Raux}. (b) Markedly different from liquid impact, cornstarch and polystyrene suspension drops were found to remain at a constant height (independent of impact velocity) for long times after impacting on a solid surface, thus providing evidence of impact-induced solidification. Adapted from~\cite{Boyer}.  (c) Impact of concentrated colloidal suspensions at conditions near to the shear thickening transition shows exotic behaviours, from the appearance of localised areas of solidification (left) to partial solidification of the whole drop (center) to near complete solidification at impact (right). Adapted from~\cite{Shah_2021_impact}.}
        
    \label{fig:shearthicken}
\end{figure*}
 
Similar to polymeric fluids, substrate wettability has a relatively small effect on the post-impact spreading of suspension drops (especially in the high-$We$ limit) \textbf{[Fig.\ \ref{fig:shearthicken}a]}, but the receding phase depends strongly on substrate properties as well as the viscosity of the suspending liquid. This asymmetry in interactions can have a strong effect on the particle distribution within the suspension after the impact process.  In particular, several works have shown that the dynamics of the receding phase are what determine the final distribution  of particles deposited on the substrate. Nicolas et al.~\cite{nicolas_2005} found that long after impact, the particle distribution on the substrate varied drastically with the suspending fluid $Re$: particles were concentrated in an annular region for impacts at high $Re$, but were uniformly distributed for low-$Re$ impacts [Fig.~\ref{fig:nonBrownian}(c)]. Grishaev et al.~\cite{GRISHAEVdist} have additionally reported that while the particles formed monolayers on hydrophilic surfaces, three-dimensional crown-like structures formed after impact on hydrophobic substrates. Thus, in addition to the viscosity of the fluid phase, substrate wettability has a direct effect on the post-impact particle distribution; understanding how to achieve a uniform particle distribution after impact is highly relevant for optimizing coating and printing applications.

As with Newtonian~\cite{arora2018smalltargets, vernay_2015_smalltargets, smalltargets_newtonian} and polymeric~\cite{Rozhkov2003} impacts, small target-based experiments have been used to minimize substrate effects in the post-impact behaviour of particulate drops. Experiments on small targets by Raux et al.\ ~\cite{Raux} showed that $d_{max}$ and $t_{max}$ after impact were independent of particle size (varied from $40$ $\mu$m to $140$ $\mu$m). However, the receding phase was found to be slowed down by the  presence of larger particles. Larger particles have also been observed to make the initially smooth film unstable during retraction, leading to rupture and decrease of film lifetime. Thus, particulate additives have a destabilizing effect on the fluid film formed after impact on small targets, in stark contrast with the stabilizing effect of polymeric additives on the film~\cite{Rozhkov2003}. This stabilizing effect of polymers has also been observed in suppressing the rebound of dilute silica suspensions~\cite{Zang2013}. To the first order, this may be understood by considering the interplay between the thickness of the spreading drop and the lengthscale of the particulate additives. As granular additives, such as studied in~\cite{Raux} are comparable in size to the spreading fluid layer, they may seed instabilities in the film. On the other hand, as polymeric additives are molecular in size (much smaller than the fluid thickness), the dynamics may be dominated by microstructural interactions.\par 

In addition to spreading, controlling the splashing of particulate fluids is key in many processes. The impact dynamics of blood, a well-known shear thinning suspension of platelet cells, are of special interest due to its relevance in forensic analyses~\cite{smith2018wetting}. De Goede et al.~\cite{deGoede2018} observed that substrate wettability had little effect on the splashing threshold of blood. This is once again consistent with observations of Newtonian splashing on substrates of varying wettability~\cite{Riboux2014}. For suspensions comprised of larger particles, where particle inertia plays a significant role, the splashing onset is fundamentally different and is best understood via particle-based parameters, as discussed in section~\ref{particle-based}.

At high concentrations (and/or high shear), particulate additives substantially modify fluid flow behaviour, most notably by inducing transient solidification.   Shear thickening fluids typically show an increasing viscosity with with increasing shear, often transitioning to  shear jamming (solid-like behaviour) at the highest stresses. Drop impact studies in the shear thickening regime result in exotic behaviours due to the large and instantaneous shear rates ($\mathcal{O}(10^{3})$ and greater), localised at the point of impact. Despite having great potential to expand our knowledge of the high stress response of these materials,
drop impact studies of shear thickening fluids have been few and far between, likely due to the experimental challenges inherent to  working with dense 
suspensions. \par

On impact, dense suspension drops have been observed to undergo shear jamming; partially or completely solidifying~\cite{Boyer, Shah_2021_impact} \textbf{[Fig.\ \ref{fig:shearthicken} b,c]}. Different studies have reported drastically different timescale over which shear jammed drops stayed solidified. Boyer et al.~\cite{Boyer} observed that shear-thickening drops (cornstarch and polystyrene suspensions, $d\sim 5$ to $20 \mu m$, volume fraction $\phi > 0.33$) showed a maximum deformation that was independent of the impact velocity, and that the drops stayed immobile long after impact [Fig.~\ref{fig:shearthicken}(b)]. Recent work by Shah et al.~\cite{Shah_2021_impact} observed shear-thickening colloidal drops ($d = 0.8 \mu$m, $\phi=0.1-0.5$) to undergo partial shear jamming, where the bottom of the drop shear jammed but the top portion deformed like a fluid [Fig.~\ref{fig:shearthicken}(c)]. At even higher impact velocities (corresponding to higher shear rates), the drop did not deform at all, and was even observed to rebound. The time evolution of drop height in this regime is reported to make a sudden transition from a `free-fall regime' (drop apex moves at the impact velocity) to a `plateau regime' (height of the drop apex constant, indicating shear jamming); this contrasts with a smoothly decreasing drop height reported in Newtonian impact~\cite{JosserandReview, ChengStressReview}.  While shear jamming observed on hydrophilic substrates was transient and the drops `unjammed' into the fluid state over a few seconds~\cite{Shah_2021_impact}, concentrated colloidal drops($\phi \sim 0.6$) impacted over hydrophobic PTFE surfaces are reported to stay jammed for days~\cite{bertola2015impact}. The interaction between the drop and the surface thus seems to play a key role in the timescale of unjamming. 

The surface of an impacting drop is deformable, enabling a unique method to observe how shear-jamming occurs in situ.  The freely deformable surface allows for direct observation to localised, macroscopic  changes in the impacting fluid drop which reflect underlying changes in suspension microstructure.  For example, shear fronts traveling upward along the drop surface at speeds much faster than $u_0$ have been reported in recent work on colloidal drop impact~\cite{Shah_2021_impact}. Observations of large frequency changes due to polystyrene drops impacting on microresonators~\cite{KIM2019QCM} have also been attributed to shear-induced structures inside the drop. Outside of drop impact, similar shear fronts have been observed in larger reservoirs of complex fluids are impacted with an impeller~\cite{waitukaitis2012impact,han2016high,peters2016direct,HanJaegerShearFronts,RomckePetersPRF2021}, and microscopic observations of local particle density modulations due to propagating shear have unveiled insights of front propagation in these systems. Similar microscopic characterisation of shear fronts in impacting drops, although challenging, would play a great role in developing the broader physics of the shear jamming transition.\par

In future work on suspension drop impact, extensive use of techniques to measure local stresses in an impacting drop is key, in order to capture the spatiotemporal variation of the stress during impact. Studies of Newtonian fluids have established the velocity and pressure fields within an impacting drop for a large range of viscosities~\cite{philippi_2016_newtonian,gordillo_force_2018}. Although these works are not directly applicable to highly non-Newtonian fluids, the experimental techniques should serve as a foundation to extend our understanding of flows inside an impacting suspensions drop. Additional challenges in data collection due to the opacity of dense particulate suspensions need to be overcome. Numerical work on the spatial variation of shear stresses during all phases of suspension drop impact could provide a necessary phenomenological basis for understanding the flow of these liquids at high stresses. \par

Current drop impact studies of shear-thickening fluids have clearly identified novel behaviours, such as partial/complete solidification upon impact.  To use drop impact as platform to study stress-induced jamming, this work should be expanded, with special attention paid to suspensions in both the shear thickening and the shear jamming limits. Overall shear stresses in impacting drops can be estimated and measured, but localised information is quite challenging obtain experimentally. To understand the physics of localised shear jamming, systematic measurements of shear stresses in the impacted drop are crucial. Promising avenues for obtaining these measurements include localised measurements of boundary stresses (akin to traction force microscopy measurements~\cite{style2014traction, Rathee_JRheol, rathee2022structure}). Shear fronts in particulate systems under confinement have been previously characterised~\cite{waitukaitis2012impact,han2016high,peters2016direct,HanJaegerShearFronts,RomckePetersPRF2021}, and numerical studies exploring how similar fronts propagate in free-surface systems would be informative. \par

Another avenue ripe for exploration is the role of particle shape. It is well-known from  bulk rheological measurements that
the critical volume fraction for jamming varies with particle shape. In particular, elongated (rod-like) particles can exhibit jamming behaviour at dramatically lower volume fractions than their spherical counterparts~\cite{williams2003, james2019c}. However, the role of particle shape on shear jamming in drop impact systems has thus far been unexplored.
These studies have the potential to probe the role of shape asymmetry in the high-stress behaviour of suspensions, as well as relevance to industrial processes, where the component particles in slurries and suspensions are often far from spherical.\par

In summary, the impact of low- and moderate-$\phi$ drops can be understood using the bulk effective viscosity of the suspension for a wide range of particle sizes. However, for high-$\phi$ drops containing larger particles, individual particle inertia plays a significant role in the impact dynamics.  In these systems, the  liquid merely acts as an agent that binds the particles into a drop, and the suspension can no longer be modelled as an effective medium. As we will discuss in the next section, particle-based parameters have shown success in characterizing the impact dynamics in this regime. \par

\subsection{Particle inertia-dominated regime}\label{particle-based} 

 \begin{figure}[h]
\centering
\includegraphics[height=13cm]{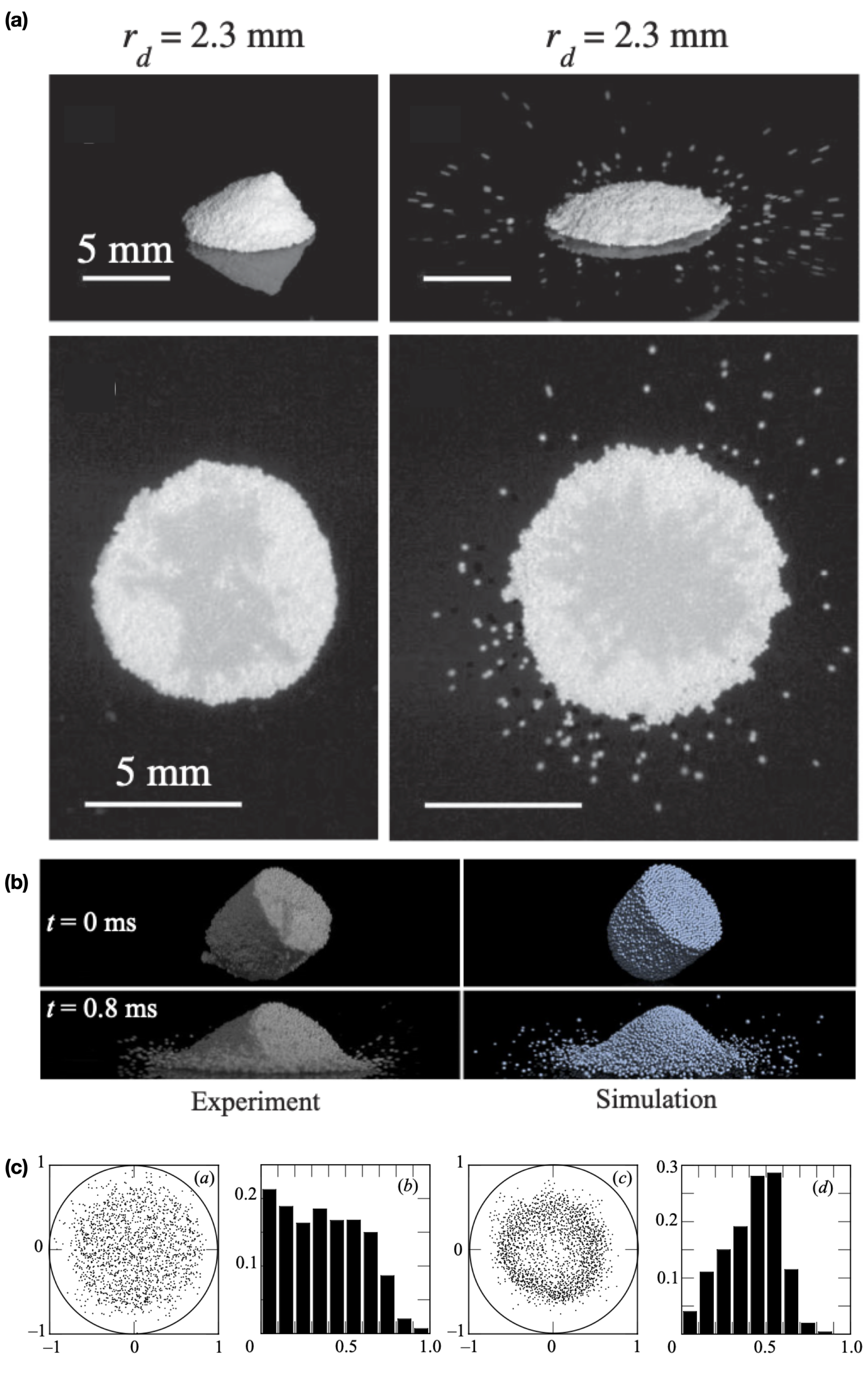}
    \caption {{Impact of granular suspensions.} (a) Upon impact, granular suspensions can either remain as a pile (splat) or spread out and eject particles (splash).  Adapted from \cite{peters2013splashing}.  (b) Molecular dynamics simulations, adapted with a force law to capture capillary interactions between particles, capture impact behaviour well in this regime. Adapted from \cite{Schaarsberg}. (c) Radial distribution of particles in the spread droplets is uniform for low Reynolds number of surrounding fluid, and annular for high $Re$. Adapted from \cite{nicolas_2005}.
}
    \label{fig:nonBrownian}
\end{figure}

`Granular suspensions', composed  of large (here greater than $\sim$ 100 microns) particles at high concentrations behave quite differently upon impact than colloidal suspensions: individual particle inertia  can play a dominant role in the physics of granular suspension impact, as opposed to bulk properties due to particle interactions.  On the other hand, granular suspensions show unique behaviours not exhibited by dry granular media due to the surrounding liquid holding the particles together~\cite{wetGranularReview}. As we discuss below, especially at high impact velocities, these systems are better understood in a particle-based manner, rather than in the bulk rheological context used to understand the impact of colloidal suspension drops.

Close to the critical volume fraction $\phi_m$ for jamming, impacting suspension drops composed of large particles have been observed to deviate from their bulk behaviour, and a clear deviation from the effective viscosity framework used to describe colloidal suspensions has been identified~\cite{jorgensen2020deformation}.  Lubbers et al.~\cite{Lubbers} studied the high-inertia impact of dense suspensions ($d=250 \mu m, \phi>0.60$). These drops created a particle monolayer post-impact, which was found to grow at a different rate than the spread of a Newtonian liquid drop on a surface. This high-$\phi$, inertia-dominated phenomenon was explained in terms of a particle-based Weber number $We_p$ and  liquid Stokes number $St$ ($We_p \gg 1$,  $St \gg 1$ in this study). The authors proposed a particle-based chain model of spreading; this emphasizes the quite different physics of impacting drops with larger particulate additives.  \par

Similar to Newtonian drops, particulate suspensions may splash under certain impact conditions. The nature of this splashing, however, is fundamentally different than that of Newtonian fluids, especially for large particle additives.  Peters et al.~\cite{Peters} studied the splashing threshold of dense non-Brownian suspensions (particle sizes greater than $80$ $\mu m$, in the range $0.59 < \phi < 0.65$). `Splashing' in this case comprised of individual particles being ejected from the edge of the drop \textbf{[Fig.\ \ref{fig:nonBrownian}a]}. They found that this splashing threshold was best characterised in terms of the particle-based Weber number $We_p=\frac{\rho_p r_p u_0^2}{\sigma}$, rather than the typical fluid $We$. This suggests that splashing occurs when an individual particle overcomes the surface energy of the surrounding liquid, and that larger and denser particles are more likely to escape at lower drop velocities. As opposed to Newtonian splashing, the onset of splashing for these suspensions (at $We_p\geq 14$) was independent of the substrate wettability and roughness. This $We_p$ dependence has been verified in subsequent studies ~\cite{Schaarsberg,Grishaev}. Consequently, Schaarsberg et al.~\cite{Schaarsberg} experimentally investigated the effect of suspending liquid viscosity on dense ($\phi=0.59$) suspension splashing, showing  that the phase space defined by $We_p$ and $St$, the Stokes number,  is cleanly divided into splashing and non-splashing regimes. Moreover, they were able to capture impact behaviour using modified molecular dynamics simulations, in which a force law between particles mimicking capillary interactions was implemented \textbf{[Fig.\ \ref{fig:nonBrownian}b]}.  Thus, the understanding has emerged that particle inertia, along with viscous interactions between particles and the suspending liquid, control the splashing threshold of dense non-Brownian suspensions, as opposed to the bulk rheological properties that govern the spreading dynamics.\par

Marston et al.~\cite{Marston} tracked individual particles ejected after splashing (grain size $\sim 350 \mu m$) and found that the maximum particle velocity was typically twice that of the drop velocity, which is much slower than ejected liquid droplet speeds in Newtonian splashing. Their image analysis suggests that the drop compressed after impact, and particles were ejected when this compression pushed $\phi$ closer to the jamming threshold. For a more comprehensive understanding of this correlation between splashing and the jamming volume fraction $\phi_m$, the elastic energy of the jammed network may need to be taken into account, as long-range correlations among the particle structure are likely to be significant near $\phi_m$.

Suspensions involved in many real-life processes are rarely monodisperse, which necessitates understanding the splashing of polydisperse suspensions. For bimodal suspensions (suspensions comprised of two particle sizes), Peters et al.~\cite{Peters} found that smaller particles were more likely to be ejected than larger particles. This is seemingly contradictory to the $We_p$ based predictions discussed above. On a close inspection, however, particle-particle collisions during impact cause smaller particles to gain higher velocities due to momentum conservation,, and thus we should expect that smaller particles get ejected earlier. This argument can potentially be extended to polydisperse suspensions, where we would expect smaller particles to eject with more ease during splashing. \par

Thus, the splashing of dense non-Brownian suspensions is governed by individual particle inertia, while the splashing of Brownian fluids is governed by bulk flow, similar to Newtonian fluids. Recently, Grishaev et al.~\cite{GRISHAEV2019_splashing} studied the splashing of suspensions made up of 10 \textmu m spheres, and found that the splashing threshold did not agree with either the bulk fluid models for Newtonian fluids~\cite{nicolas_2005} or particle-based models ~\cite{Grishaev,Peters}. The droplets ejected after splashing were an order of magnitude larger than the particulate additives, as opposed to individual particles ejecting for suspensions whose splashing was governed by particle-based parameters~\cite{Grishaev,Peters}. This indicates a broad crossover regime may exist  between the two extremes of suspension splashing --- a regime that is governed by bulk flow and one that is governed by individual particle inertia. As studying individual particle dynamics at such short timescales requires very high spatiotemporal resolution, more experimental studies that integrate microscopy with high-speed imaging might be fruitful in this respect. Although specific impact regimes have been studied in detail, a large amount of systematic data ranging over particle Weber numbers is necessary to develop scaling laws over the whole range of particle sizes. \par

In addition to size heterogeneity, exploration of the how particle shape modifies impact outcomes is crucial for applications development; nearly all industrial suspensions are irregularly shaped and polydisperse. As non-spherical particles additionally introduce another degree of freedom in particle behaviour, this line of investigation may uncover a plethora of rich physical phenomena.  Particle shape affects not only the the critical volume fraction for jamming, $\phi_m$~\cite{williams2003, james2019c}, but additionally the ability of particles to align due to the large shear present in a spreading drop.  The realignment of particles due to shear can modify bulk flow properties of the suspension, and is highly likely to modify impact behaviour. In addition to exploring the role of particle shape, future work should move towards direct measurements of suspension microstructure and flow, so that it may directly connected to impact outcomes.  In the dense suspension limit where the bulk viscosity framework is inadequate, experiments with index-matched fluids with tracer particles, though extremely challenging,  would be worthwhile to visualize microstructure in the drop drop during impact. This regime is particularly challenging to explore via simulations, so obtaining highly spatially resolved experimental  data for $\phi$ is crucial to enable progress in establishing constitutive equations for these complex materials.

\section{Outlook} \label{conclusion}

In this review, we have summarized current work on the drop impact of complex fluids on solid substrates. Drop impact enables us to study fluid properties at very high stresses, often beyond the range of typical rheometers, and in the presence of a free deformable surface.
Insights developed in this field are not only limited to  drop impact, but have tremendous potential to further our understanding of the flow behaviour of complex fluids under dynamic conditions. \par

Understanding and controlling complex fluid behaviour is crucial to efficiency and performance in many industries, for example food, personal and home care, and additive manufacturing. Newtonian drops, specifically water,  impacting a solid or liquid substrate play an important role in a multitude of processes such as geological erosion, mixing of air into oceans, and wear of turbines. Droplet impact of complex fluids features in many industrial and natural processes. For example, colloidal droplet impact is relevant to optimizing inkjet printing~\cite{lohse2022fundamental} and performing forensic analyses~\cite{smith2018wetting}. Pesticides and insecticides sprayed on crops are polymeric in nature, and this modifies their impact dynamics. Airborne droplets that spread infectious diseases contain microorganisms of colloidal size, altering droplet breakup and the spread of infections. Industrial coatings contain both polymeric and colloidal components~\cite{breitenbach2018drop}. Coatings are sprayed onto substrates to impart them with properties such as durability, a matte or shiny appearance, and hydrophobicity. Pharmaceutical tablets are also spray coated with polymeric fluids for a number of functions such as masking of odors and taste, or slow release of medication into the bloodstream~\cite{bolleddula2010impact}. Impact of Newtonian or non-Newtonian droplets on complex substrates pertains to processes such as spray cooling of hot surfaces and raindrops mixing with porous and granular media such as soil or sand. Thus, a foundational understanding of the physics of droplet impact is important to ensure superior and consistent quality of a plethora of products. 

While there has been extensive work on the impact of Newtonian drops, non-Newtonian fluid impact is a relatively young field. Due to and highly multidisciplinary nature  and variety of applications of this process, studies of impacting complex fluids have so far followed disparate avenues, and few broad insights have thus far been drawn. The rich rheological properties of non-Newtonian fluids may manifest in many ways under short timescales and free-surface conditions present during impact, but traditional rheometry techniques can only probe fluid behaviour under steady strain and confined conditions.  Thus, information gathered from standard rheometry is inadequate to build a complete understanding of complex fluids under dynamic conditions. Drop impact provides an ideal way to investigate how complex rheology may manifest in a freely deformable surface under high localised shear experienced at the impact point.

In this review, we have attempted to classify results in non-Newtonian drop impact based on the two broadly encountered material compositions: polymeric fluids and particulate suspensions. We have presented a fluid dynamics and rheology-based treatment of the current understanding of complex fluid drop impact. We hope that this work serves as a reference point across disciplinary boundaries, and helps move towards a holistic understanding of the physics of complex fluids under high stresses and free surface conditions.

Both polymeric and particulate fluids show shear thinning within specific parameter ranges. Surprisingly, for both these categories, post-impact spreading in the shear-thinning regime is successfully modeled using the effective fluid viscosity and existing Newtonian models~\cite{AnLee2012, Guemas,German_2009, Shah_2021_impact,jorgensen2020deformation,Ok,nicolas_2005,Boyer, Raux, Marston, Grishaev, GRISHAEVdist}. However, in other parameter regimes, a number of exotic impact behaviours have been reported for complex fluids that have no counterparts in Newtonian impact. 

Polymer additives suppress droplet bounce when impacted on hydrophobic surfaces~\cite{Bergeron2000,Bartolo2007,SmithBertola2010,Zang2013,CrooksBoger2000}. This behaviour cannot be satisfactorily explained via bulk fluid properties such as elongational viscosity or normal stresses; contact line interactions of polymer molecules are emerging as a mechanism behind bounce suppression~\cite{BERTOLA2013polymerimpact}. Dense suspensions undergo shear thickening at high stresses, and many novel behaviours including partial and complete solidification have been reported in this regime~\cite{Boyer, Shah_2021_impact, bertola2015impact}. Dense suspensions with larger particle additives spread and splash in a fundamentally different manner than Newtonian and colloidal fluids, and their dynamics must be characterised using particle-based models~\cite{Lubbers, Peters, Schaarsberg, Grishaev} (as opposed to bulk suspension properties). Thus depending on the parameter regime, specifically in additive concentration and applied shear, either bulk properties or localised dynamics may control the impact outcomes.  

In specific regimes, scaling laws have successfully characterised impact dynamics, and some numerical studies have explored the effect of each fluid property on impact behaviour. However, a more unified description of the physics of complex fluid impact still eludes us. Below, we highlight prominent future directions necessary to build a comprehensive understanding of non-Newtonian drop impact. We draw connections with complex fluid studies in fields adjacent to drop impact (for example, bulk rheology~\cite{StickelPowellRheo,DennMortonSoftMatter} and impeller impact on complex fluids under confinement~\cite{peters2016direct, RomckePetersPRF2021, han2016high}) that are key to building a unified description of the physics of non-Newtonian fluids. 

In order to develop a more detailed understanding of complex fluid drop impact, systematic data spanning a large range of impact $We$ and $Re$ needs to be collected. Detailed experiments especially in the highly non-Newtonian regimes would provide a window into the more exotic impact outcomes observed. We now have the ability to record the macroscopic details of impacting drops at higher speeds than ever before. Future experiments should combine high-speed imaging with other tools such as Particle Image Velocimetry~\cite{westerweel2013particle} and local stress measurements~\cite{ChengStressReview, Rathee_JRheol} to develop insights on both bulk and microscopic levels. Extensive data collected from multiple channels would enable us to build a physical understanding over the entire parameter space, and also control droplet behaviours is various applications.


To date, studies on particulate drops have not systematically explored the effect of particle shape. 
The most convenient avenue to investigate the role of particle anisotropy on impact dynamics is by varying particle aspect ratio.  Using elongated particles alters suspension behaviour in two major ways: Bulk rheological studies of rod-shaped particle suspensions show that both the threshold and range of shear thickening are highly modified due to particle shape~\cite{williams2003, james2019c}. The broader range of shear thickening in anisotropic suspensions provides a larger state space to explore exotic behaviours such as solidification and bounce upon impact. Additionally, the shape of particle additives may significantly modify contact-line dynamics. Numerous studies have reported that for both particulate and polymeric fluids, dynamics at the contact line can play a significant role in impact behaviours. Thus, the impact of suspension droplets with elongated particles is an ideal system to unify existing insights on contact-line dynamics in these two classes of fluids. Additionally, the effect of polymer architecture, especially branching, on impact behaviour  needs to be paid more attention in future experiments on polymeric fluids.

Our understanding of the behaviour of impacting complex fluids is inevitably built on the current understanding of the rheology of these fluids. Although a large amount of bulk rheological data is available and new data is being added over a range of timescales and shear rates, developing constitutive models for non-Newtonian flow behaviour is still a highly active avenue of research~\cite{StickelPowellRheo}. 
Measurements of localised stress and deformation in future work would directly apply to free-surface systems such as drop impact, and complement data available from bulk rheology. We hope that systematic drop impact experiments  equipped with local measurements could potentially inform further work on constitutive rheological models especially at high shear stresses, allowing these disciplines to co-evolve in the near future.  \par

The focus of this review was impacting drops on smooth, rigid, dry substrates. However, in many real-life processes, impact occurs on rough, wet, curved, compliant, or heated surfaces. While Newtonian impact studies have begun to focus on these aspects, exploration of the impact behaviour of complex fluids on such substrates is limited. Some experimental works have dealt with complex fluid impact on substrates thinly coated with liquid~\cite{Blackwell2015, sen_morales_ewoldt_2020, tang2019spreading}, and developed an understanding of impact in specific parameter regimes. Other work has explored impact of polymeric fluids on heated substrates~\cite{bertola2012bookchapter,BERTOLA2014259}, showing that even small amounts of polymer additives decreases the dynamic Leidenfrost temperature of water, and inhibits droplet splashing and atomization when impacted on heated substrates. Heating of substrates has been reported to modify deposition patterns of sessile colloidal drops~\cite{patil2016effects}, and even a small amount of particles have been shown to modify drop impact dynamics in both boiling and Leidenfrost regimes~\cite{ma2024nanofluid}. Hydrogels on heated substrates `harvest' energy from the substrates to sustain bouncing~\cite{waitukaitis2017coupling, pham2017spontaneous}, with significant implications for the emerging field of soft robotics. There could be overlap in the underlying mechanism for polymeric fluids and hydrogel impact, such as elasticity imparted by polymeric additives. Exploration of modified substrate properties on impacting complex fluid drops is necessary to connect with applications. Numerical work that varies the mechanical properties of substrates along with the fluid characteristics might be beneficial in this respect. Development of experimental techniques that record both macroscopic and microscopic behaviours will also benefit further work.  

A large number of drops impact a substrate at close distances in spray applications. The effect of neighbouring droplets on spreading and receding dynamics are an important phenomenon that has not yet received much attention. While most drop impact studies focus on millimetric droplets, applications involving aerosols and sprays pertain to microdroplets. Droplet size heavily modifies dimensionless parameters such as $Re$ and $We$, and the ratio of the lengthscale of particulate/polymeric additive to drop size is also key to droplet behaviour after impact. Many industrial fluids contain highly evaporative components, e.g. alcohol-based solvents. Evaporation effects on impact are even more prominent for small droplet sizes. Therefore, future work focusing on microdroplet impact and the simultaneous impact of multiple droplets would be of high industrial relevance. 

For impacting complex fluids, interaction lengthscales and  timescales between the the solid and liquid phases compete with the already existing scales such as drop size, additive size, surface roughness., and thixotropic recovery. While such a large variety of competing lengthscales and timescales enriches the drop impact problem, it also makes it extremely challenging. The availability of a number of microscopy techniques, and advances in high-speed imaging and data storage capacity make this an exciting time to attack this problem. The time is ripe for widespread use of these techniques to obtain more spatio-temporally resolved data, laying the groundwork for a deeper theoretical understanding of complex fluid behaviour.

\section*{Author Contributions}
PS and MMD drafted the manuscript.

\section*{Conflicts of interest}
There are no conflicts to declare.

We thank Jeff Richards, Xiang Cheng, Brendan Blackwell, and Brennan Sprinkle for fruitful discussions and comments.

\bibliography{drop_impact} 

\providecommand*{\mcitethebibliography}{\thebibliography}
\csname @ifundefined\endcsname{endmcitethebibliography}
{\let\endmcitethebibliography\endthebibliography}{}
\begin{mcitethebibliography}{144}
\providecommand*{\natexlab}[1]{#1}
\providecommand*{\mciteSetBstSublistMode}[1]{}
\providecommand*{\mciteSetBstMaxWidthForm}[2]{}
\providecommand*{\mciteBstWouldAddEndPuncttrue}
  {\def\EndOfBibitem{\unskip.}}
\providecommand*{\mciteBstWouldAddEndPunctfalse}
  {\let\EndOfBibitem\relax}
\providecommand*{\mciteSetBstMidEndSepPunct}[3]{}
\providecommand*{\mciteSetBstSublistLabelBeginEnd}[3]{}
\providecommand*{\EndOfBibitem}{}
\mciteSetBstSublistMode{f}
\mciteSetBstMaxWidthForm{subitem}
{(\emph{\alph{mcitesubitemcount}})}
\mciteSetBstSublistLabelBeginEnd{\mcitemaxwidthsubitemform\space}
{\relax}{\relax}

\bibitem[Yarin(2006)]{YarinReview}
A.~Yarin, \emph{Annual Review of Fluid Mechanics}, 2006, \textbf{38},
  159--192\relax
\mciteBstWouldAddEndPuncttrue
\mciteSetBstMidEndSepPunct{\mcitedefaultmidpunct}
{\mcitedefaultendpunct}{\mcitedefaultseppunct}\relax
\EndOfBibitem
\bibitem[Josserand and Thoroddsen(2016)]{JosserandReview}
C.~Josserand and S.~Thoroddsen, \emph{Annual Review of Fluid Mechanics}, 2016,
  \textbf{48}, 365--391\relax
\mciteBstWouldAddEndPuncttrue
\mciteSetBstMidEndSepPunct{\mcitedefaultmidpunct}
{\mcitedefaultendpunct}{\mcitedefaultseppunct}\relax
\EndOfBibitem
\bibitem[Worthington(1908)]{Worthington}
A.~M. Worthington, \emph{London: Longmans, Green}, 1908\relax
\mciteBstWouldAddEndPuncttrue
\mciteSetBstMidEndSepPunct{\mcitedefaultmidpunct}
{\mcitedefaultendpunct}{\mcitedefaultseppunct}\relax
\EndOfBibitem
\bibitem[Xu \emph{et~al.}(2005)Xu, Zhang, and Nagel]{Xu2005}
L.~Xu, W.~W. Zhang and S.~R. Nagel, \emph{Physical review letters}, 2005,
  \textbf{94}, 184505\relax
\mciteBstWouldAddEndPuncttrue
\mciteSetBstMidEndSepPunct{\mcitedefaultmidpunct}
{\mcitedefaultendpunct}{\mcitedefaultseppunct}\relax
\EndOfBibitem
\bibitem[Breitenbach \emph{et~al.}(2018)Breitenbach, Roisman, and
  Tropea]{breitenbach2018drop}
J.~Breitenbach, I.~V. Roisman and C.~Tropea, \emph{Experiments in Fluids},
  2018, \textbf{59}, 1--21\relax
\mciteBstWouldAddEndPuncttrue
\mciteSetBstMidEndSepPunct{\mcitedefaultmidpunct}
{\mcitedefaultendpunct}{\mcitedefaultseppunct}\relax
\EndOfBibitem
\bibitem[Lohse(2022)]{lohse2022fundamental}
D.~Lohse, \emph{Annual review of fluid mechanics}, 2022, \textbf{54},
  349--382\relax
\mciteBstWouldAddEndPuncttrue
\mciteSetBstMidEndSepPunct{\mcitedefaultmidpunct}
{\mcitedefaultendpunct}{\mcitedefaultseppunct}\relax
\EndOfBibitem
\bibitem[Smith and Brutin(2018)]{smith2018wetting}
F.~Smith and D.~Brutin, \emph{Current Opinion in Colloid \& Interface Science},
  2018, \textbf{36}, 78--83\relax
\mciteBstWouldAddEndPuncttrue
\mciteSetBstMidEndSepPunct{\mcitedefaultmidpunct}
{\mcitedefaultendpunct}{\mcitedefaultseppunct}\relax
\EndOfBibitem
\bibitem[Bolleddula \emph{et~al.}(2010)Bolleddula, Berchielli, and
  Aliseda]{bolleddula2010impact}
D.~Bolleddula, A.~Berchielli and A.~Aliseda, \emph{Advances in colloid and
  interface science}, 2010, \textbf{159}, 144--159\relax
\mciteBstWouldAddEndPuncttrue
\mciteSetBstMidEndSepPunct{\mcitedefaultmidpunct}
{\mcitedefaultendpunct}{\mcitedefaultseppunct}\relax
\EndOfBibitem
\bibitem[Shah \emph{et~al.}(2022)Shah, Arora, and Driscoll]{Shah_2021_impact}
P.~Shah, S.~Arora and M.~M. Driscoll, \emph{Communications Physics}, 2022,
  \textbf{5}, 1--9\relax
\mciteBstWouldAddEndPuncttrue
\mciteSetBstMidEndSepPunct{\mcitedefaultmidpunct}
{\mcitedefaultendpunct}{\mcitedefaultseppunct}\relax
\EndOfBibitem
\bibitem[Marston \emph{et~al.}(2013)Marston, Mansoor, and Thoroddsen]{Marston}
J.~O. Marston, M.~M. Mansoor and S.~T. Thoroddsen, \emph{Phys. Rev. E}, 2013,
  \textbf{88}, 010201\relax
\mciteBstWouldAddEndPuncttrue
\mciteSetBstMidEndSepPunct{\mcitedefaultmidpunct}
{\mcitedefaultendpunct}{\mcitedefaultseppunct}\relax
\EndOfBibitem
\bibitem[Blackwell \emph{et~al.}(2015)Blackwell, Deetjen, Gaudio, and
  Ewoldt]{Blackwell2015}
B.~C. Blackwell, M.~E. Deetjen, J.~E. Gaudio and R.~H. Ewoldt, \emph{Physics of
  Fluids}, 2015, \textbf{27}, 043101\relax
\mciteBstWouldAddEndPuncttrue
\mciteSetBstMidEndSepPunct{\mcitedefaultmidpunct}
{\mcitedefaultendpunct}{\mcitedefaultseppunct}\relax
\EndOfBibitem
\bibitem[Bertola \emph{et~al.}(2012)Bertola, Marengo, Ferrari, Liggieri, and
  Miller]{bertola2012bookchapter}
V.~Bertola, M.~Marengo, M.~Ferrari, L.~Liggieri and R.~Miller, \emph{Drops and
  Bubbles in Contact with Solid Surfaces}, 2012,  267--298\relax
\mciteBstWouldAddEndPuncttrue
\mciteSetBstMidEndSepPunct{\mcitedefaultmidpunct}
{\mcitedefaultendpunct}{\mcitedefaultseppunct}\relax
\EndOfBibitem
\bibitem[Aksoy \emph{et~al.}(2022)Aksoy, Eneren, Koos, and
  Vetrano]{AKSOY2022nanofluid}
Y.~Aksoy, P.~Eneren, E.~Koos and M.~Vetrano, \emph{Journal of Colloid and
  Interface Science}, 2022, \textbf{606}, 434--443\relax
\mciteBstWouldAddEndPuncttrue
\mciteSetBstMidEndSepPunct{\mcitedefaultmidpunct}
{\mcitedefaultendpunct}{\mcitedefaultseppunct}\relax
\EndOfBibitem
\bibitem[Mewis and Wagner(2012)]{Wagner2012colloidal}
J.~Mewis and N.~J. Wagner, \emph{Colloidal suspension rheology}, Cambridge
  University Press, 2012\relax
\mciteBstWouldAddEndPuncttrue
\mciteSetBstMidEndSepPunct{\mcitedefaultmidpunct}
{\mcitedefaultendpunct}{\mcitedefaultseppunct}\relax
\EndOfBibitem
\bibitem[Shaw(2012)]{shaw2012introduction}
M.~T. Shaw, \emph{Introduction to polymer rheology}, John Wiley \& Sons,
  2012\relax
\mciteBstWouldAddEndPuncttrue
\mciteSetBstMidEndSepPunct{\mcitedefaultmidpunct}
{\mcitedefaultendpunct}{\mcitedefaultseppunct}\relax
\EndOfBibitem
\bibitem[Macosko(1994)]{macosko1994rheology}
C.~W. Macosko, \emph{Measurements and Applications}, 1994\relax
\mciteBstWouldAddEndPuncttrue
\mciteSetBstMidEndSepPunct{\mcitedefaultmidpunct}
{\mcitedefaultendpunct}{\mcitedefaultseppunct}\relax
\EndOfBibitem
\bibitem[Denn \emph{et~al.}(2018)Denn, Morris, and Bonn]{DennMortonSoftMatter}
M.~M. Denn, J.~F. Morris and D.~Bonn, \emph{Soft Matter}, 2018, \textbf{14},
  170--184\relax
\mciteBstWouldAddEndPuncttrue
\mciteSetBstMidEndSepPunct{\mcitedefaultmidpunct}
{\mcitedefaultendpunct}{\mcitedefaultseppunct}\relax
\EndOfBibitem
\bibitem[Osswald and Rudolph(2015)]{osswald2015polymer}
T.~Osswald and N.~Rudolph, \emph{Polymer rheology}, 2015\relax
\mciteBstWouldAddEndPuncttrue
\mciteSetBstMidEndSepPunct{\mcitedefaultmidpunct}
{\mcitedefaultendpunct}{\mcitedefaultseppunct}\relax
\EndOfBibitem
\bibitem[Barnes(1999)]{BARNES1999133}
H.~A. Barnes, \emph{Journal of Non-Newtonian Fluid Mechanics}, 1999,
  \textbf{81}, 133 -- 178\relax
\mciteBstWouldAddEndPuncttrue
\mciteSetBstMidEndSepPunct{\mcitedefaultmidpunct}
{\mcitedefaultendpunct}{\mcitedefaultseppunct}\relax
\EndOfBibitem
\bibitem[Balmforth \emph{et~al.}(2014)Balmforth, Frigaard, and
  Ovarlez]{Balmforth2014}
N.~J. Balmforth, I.~A. Frigaard and G.~Ovarlez, \emph{Annual Review of Fluid
  Mechanics}, 2014, \textbf{46}, 121--146\relax
\mciteBstWouldAddEndPuncttrue
\mciteSetBstMidEndSepPunct{\mcitedefaultmidpunct}
{\mcitedefaultendpunct}{\mcitedefaultseppunct}\relax
\EndOfBibitem
\bibitem[Coussot(2014)]{COUSSOT201431}
P.~Coussot, \emph{Journal of Non-Newtonian Fluid Mechanics}, 2014,
  \textbf{211}, 31 -- 49\relax
\mciteBstWouldAddEndPuncttrue
\mciteSetBstMidEndSepPunct{\mcitedefaultmidpunct}
{\mcitedefaultendpunct}{\mcitedefaultseppunct}\relax
\EndOfBibitem
\bibitem[Herschel and Bulkley(1926)]{Herschel1926}
W.~H. Herschel and R.~Bulkley, \emph{Kolloid-Zeitschrift}, 1926, \textbf{39},
  291--300\relax
\mciteBstWouldAddEndPuncttrue
\mciteSetBstMidEndSepPunct{\mcitedefaultmidpunct}
{\mcitedefaultendpunct}{\mcitedefaultseppunct}\relax
\EndOfBibitem
\bibitem[Petrie(2006)]{extensional_PETRIE20061}
C.~J. Petrie, \emph{Journal of Non-Newtonian Fluid Mechanics}, 2006,
  \textbf{137}, 1 -- 14\relax
\mciteBstWouldAddEndPuncttrue
\mciteSetBstMidEndSepPunct{\mcitedefaultmidpunct}
{\mcitedefaultendpunct}{\mcitedefaultseppunct}\relax
\EndOfBibitem
\bibitem[Barnes(1997)]{thixotropy_BARNES19971}
H.~A. Barnes, \emph{Journal of Non-Newtonian Fluid Mechanics}, 1997,
  \textbf{70}, 1 -- 33\relax
\mciteBstWouldAddEndPuncttrue
\mciteSetBstMidEndSepPunct{\mcitedefaultmidpunct}
{\mcitedefaultendpunct}{\mcitedefaultseppunct}\relax
\EndOfBibitem
\bibitem[Mewis and Wagner(2009)]{MEWISthixotropy}
J.~Mewis and N.~J. Wagner, \emph{Advances in Colloid and Interface Science},
  2009, \textbf{147-148}, 214--227\relax
\mciteBstWouldAddEndPuncttrue
\mciteSetBstMidEndSepPunct{\mcitedefaultmidpunct}
{\mcitedefaultendpunct}{\mcitedefaultseppunct}\relax
\EndOfBibitem
\bibitem[Larson and Wei(2019)]{ThixotropyLarson}
R.~G. Larson and Y.~Wei, \emph{Journal of Rheology}, 2019, \textbf{63},
  477--501\relax
\mciteBstWouldAddEndPuncttrue
\mciteSetBstMidEndSepPunct{\mcitedefaultmidpunct}
{\mcitedefaultendpunct}{\mcitedefaultseppunct}\relax
\EndOfBibitem
\bibitem[Wyart and Cates(2014)]{WyartCates2014}
M.~Wyart and M.~E. Cates, \emph{Phys. Rev. Lett.}, 2014, \textbf{112},
  098302\relax
\mciteBstWouldAddEndPuncttrue
\mciteSetBstMidEndSepPunct{\mcitedefaultmidpunct}
{\mcitedefaultendpunct}{\mcitedefaultseppunct}\relax
\EndOfBibitem
\bibitem[Krieger and Dougherty(1959)]{krieger1959mechanism}
I.~M. Krieger and T.~J. Dougherty, \emph{Transactions of the Society of
  Rheology}, 1959, \textbf{3}, 137--152\relax
\mciteBstWouldAddEndPuncttrue
\mciteSetBstMidEndSepPunct{\mcitedefaultmidpunct}
{\mcitedefaultendpunct}{\mcitedefaultseppunct}\relax
\EndOfBibitem
\bibitem[Josserand and Zaleski(2003)]{josserand2003droplet}
C.~Josserand and S.~Zaleski, \emph{Physics of fluids}, 2003, \textbf{15},
  1650--1657\relax
\mciteBstWouldAddEndPuncttrue
\mciteSetBstMidEndSepPunct{\mcitedefaultmidpunct}
{\mcitedefaultendpunct}{\mcitedefaultseppunct}\relax
\EndOfBibitem
\bibitem[Sen \emph{et~al.}(2020)Sen, Morales, and
  Ewoldt]{sen_morales_ewoldt_2020}
S.~Sen, A.~G. Morales and R.~H. Ewoldt, \emph{Journal of Fluid Mechanics},
  2020, \textbf{891}, A27\relax
\mciteBstWouldAddEndPuncttrue
\mciteSetBstMidEndSepPunct{\mcitedefaultmidpunct}
{\mcitedefaultendpunct}{\mcitedefaultseppunct}\relax
\EndOfBibitem
\bibitem[Howland \emph{et~al.}(2016)Howland, Antkowiak, Castrej{\'o}n-Pita,
  Howison, Oliver, Style, and Castrej{\'o}n-Pita]{howland2016s}
C.~J. Howland, A.~Antkowiak, J.~R. Castrej{\'o}n-Pita, S.~D. Howison, J.~M.
  Oliver, R.~W. Style and A.~A. Castrej{\'o}n-Pita, \emph{Physical review
  letters}, 2016, \textbf{117}, 184502\relax
\mciteBstWouldAddEndPuncttrue
\mciteSetBstMidEndSepPunct{\mcitedefaultmidpunct}
{\mcitedefaultendpunct}{\mcitedefaultseppunct}\relax
\EndOfBibitem
\bibitem[Pepper \emph{et~al.}(2008)Pepper, Courbin, and Stone]{pepper2008}
R.~E. Pepper, L.~Courbin and H.~A. Stone, \emph{Physics of Fluids}, 2008,
  \textbf{20}, 082103\relax
\mciteBstWouldAddEndPuncttrue
\mciteSetBstMidEndSepPunct{\mcitedefaultmidpunct}
{\mcitedefaultendpunct}{\mcitedefaultseppunct}\relax
\EndOfBibitem
\bibitem[Bertola(2014)]{BERTOLA2014259}
V.~Bertola, \emph{Experimental Thermal and Fluid Science}, 2014, \textbf{52},
  259--269\relax
\mciteBstWouldAddEndPuncttrue
\mciteSetBstMidEndSepPunct{\mcitedefaultmidpunct}
{\mcitedefaultendpunct}{\mcitedefaultseppunct}\relax
\EndOfBibitem
\bibitem[Zhao \emph{et~al.}(2017)Zhao, de~Jong, and van~der Meer]{ZhaoPRL2017}
S.-C. Zhao, R.~de~Jong and D.~van~der Meer, \emph{Phys. Rev. Lett.}, 2017,
  \textbf{118}, 054502\relax
\mciteBstWouldAddEndPuncttrue
\mciteSetBstMidEndSepPunct{\mcitedefaultmidpunct}
{\mcitedefaultendpunct}{\mcitedefaultseppunct}\relax
\EndOfBibitem
\bibitem[GILET and BUSH(2009)]{GILET_BUSH_2009}
T.~GILET and J.~W.~M. BUSH, \emph{Journal of Fluid Mechanics}, 2009,
  \textbf{625}, 167–203\relax
\mciteBstWouldAddEndPuncttrue
\mciteSetBstMidEndSepPunct{\mcitedefaultmidpunct}
{\mcitedefaultendpunct}{\mcitedefaultseppunct}\relax
\EndOfBibitem
\bibitem[Mohammad~Karim(2023)]{KarimReview10.1063/5.0130043}
A.~Mohammad~Karim, \emph{Journal of Applied Physics}, 2023, \textbf{133},
  030701\relax
\mciteBstWouldAddEndPuncttrue
\mciteSetBstMidEndSepPunct{\mcitedefaultmidpunct}
{\mcitedefaultendpunct}{\mcitedefaultseppunct}\relax
\EndOfBibitem
\bibitem[Laan \emph{et~al.}(2014)Laan, de~Bruin, Bartolo, Josserand, and
  Bonn]{laan2014maximum}
N.~Laan, K.~G. de~Bruin, D.~Bartolo, C.~Josserand and D.~Bonn, \emph{Physical
  Review Applied}, 2014, \textbf{2}, 044018\relax
\mciteBstWouldAddEndPuncttrue
\mciteSetBstMidEndSepPunct{\mcitedefaultmidpunct}
{\mcitedefaultendpunct}{\mcitedefaultseppunct}\relax
\EndOfBibitem
\bibitem[Lee \emph{et~al.}(2016)Lee, Laan, de~Bruin, Skantzaris, Shahidzadeh,
  Derome, Carmeliet, and Bonn]{lee_laan_bonn_2016}
J.~B. Lee, N.~Laan, K.~G. de~Bruin, G.~Skantzaris, N.~Shahidzadeh, D.~Derome,
  J.~Carmeliet and D.~Bonn, \emph{Journal of Fluid Mechanics}, 2016,
  \textbf{786}, R4\relax
\mciteBstWouldAddEndPuncttrue
\mciteSetBstMidEndSepPunct{\mcitedefaultmidpunct}
{\mcitedefaultendpunct}{\mcitedefaultseppunct}\relax
\EndOfBibitem
\bibitem[Bartolo \emph{et~al.}(2005)Bartolo, Josserand, and
  Bonn]{bartolo2005retraction}
D.~Bartolo, C.~Josserand and D.~Bonn, \emph{Journal of Fluid Mechanics}, 2005,
  \textbf{545}, 329--338\relax
\mciteBstWouldAddEndPuncttrue
\mciteSetBstMidEndSepPunct{\mcitedefaultmidpunct}
{\mcitedefaultendpunct}{\mcitedefaultseppunct}\relax
\EndOfBibitem
\bibitem[Richard \emph{et~al.}(2002)Richard, Clanet, and Qu{\'e}r{\'e}]{Quere}
D.~Richard, C.~Clanet and D.~Qu{\'e}r{\'e}, \emph{Nature}, 2002, \textbf{417},
  811--811\relax
\mciteBstWouldAddEndPuncttrue
\mciteSetBstMidEndSepPunct{\mcitedefaultmidpunct}
{\mcitedefaultendpunct}{\mcitedefaultseppunct}\relax
\EndOfBibitem
\bibitem[Bird \emph{et~al.}(2013)Bird, Dhiman, Kwon, and
  Varanasi]{Bird2013contact}
J.~C. Bird, R.~Dhiman, H.-M. Kwon and K.~K. Varanasi, \emph{Nature}, 2013,
  \textbf{503}, 385--388\relax
\mciteBstWouldAddEndPuncttrue
\mciteSetBstMidEndSepPunct{\mcitedefaultmidpunct}
{\mcitedefaultendpunct}{\mcitedefaultseppunct}\relax
\EndOfBibitem
\bibitem[Riboux and Gordillo(2014)]{Riboux2014}
G.~Riboux and J.~M. Gordillo, \emph{Phys. Rev. Lett.}, 2014, \textbf{113},
  024507\relax
\mciteBstWouldAddEndPuncttrue
\mciteSetBstMidEndSepPunct{\mcitedefaultmidpunct}
{\mcitedefaultendpunct}{\mcitedefaultseppunct}\relax
\EndOfBibitem
\bibitem[Eggers \emph{et~al.}(2010)Eggers, Fontelos, Josserand, and
  Zaleski]{Eggers2010}
J.~Eggers, M.~A. Fontelos, C.~Josserand and S.~Zaleski, \emph{Physics of
  Fluids}, 2010, \textbf{22}, 062101\relax
\mciteBstWouldAddEndPuncttrue
\mciteSetBstMidEndSepPunct{\mcitedefaultmidpunct}
{\mcitedefaultendpunct}{\mcitedefaultseppunct}\relax
\EndOfBibitem
\bibitem[Lagubeau \emph{et~al.}(2012)Lagubeau, Fontelos, Josserand, Maurel,
  Pagneux, and Petitjeans]{lagubeau2012spreading}
G.~Lagubeau, M.~A. Fontelos, C.~Josserand, A.~Maurel, V.~Pagneux and
  P.~Petitjeans, \emph{Journal of Fluid Mechanics}, 2012, \textbf{713},
  50\relax
\mciteBstWouldAddEndPuncttrue
\mciteSetBstMidEndSepPunct{\mcitedefaultmidpunct}
{\mcitedefaultendpunct}{\mcitedefaultseppunct}\relax
\EndOfBibitem
\bibitem[Du \emph{et~al.}(2021)Du, Zhang, and
  Min]{DuMin2021NumericalRetraction}
J.~Du, Y.~Zhang and Q.~Min, \emph{Colloids and Surfaces A: Physicochemical and
  Engineering Aspects}, 2021, \textbf{609}, 125649\relax
\mciteBstWouldAddEndPuncttrue
\mciteSetBstMidEndSepPunct{\mcitedefaultmidpunct}
{\mcitedefaultendpunct}{\mcitedefaultseppunct}\relax
\EndOfBibitem
\bibitem[Wang and Fang(2020)]{wang2020retraction}
F.~Wang and T.~Fang, \emph{Physical Review Fluids}, 2020, \textbf{5},
  033604\relax
\mciteBstWouldAddEndPuncttrue
\mciteSetBstMidEndSepPunct{\mcitedefaultmidpunct}
{\mcitedefaultendpunct}{\mcitedefaultseppunct}\relax
\EndOfBibitem
\bibitem[Arora \emph{et~al.}(2018)Arora, Fromental, Mora, Phou, Ramos, and
  Ligoure]{arora2018smalltargets}
S.~Arora, J.-M. Fromental, S.~Mora, T.~Phou, L.~Ramos and C.~Ligoure,
  \emph{Physical Review Letters}, 2018, \textbf{120}, 148003\relax
\mciteBstWouldAddEndPuncttrue
\mciteSetBstMidEndSepPunct{\mcitedefaultmidpunct}
{\mcitedefaultendpunct}{\mcitedefaultseppunct}\relax
\EndOfBibitem
\bibitem[Vernay \emph{et~al.}(2015)Vernay, Ramos, and
  Ligoure]{vernay_2015_smalltargets}
C.~Vernay, L.~Ramos and C.~Ligoure, \emph{Journal of Fluid Mechanics}, 2015,
  \textbf{764}, 428–444\relax
\mciteBstWouldAddEndPuncttrue
\mciteSetBstMidEndSepPunct{\mcitedefaultmidpunct}
{\mcitedefaultendpunct}{\mcitedefaultseppunct}\relax
\EndOfBibitem
\bibitem[Rozhkov \emph{et~al.}(2002)Rozhkov, Prunet-Foch, and
  Vignes-Adler]{smalltargets_newtonian}
A.~Rozhkov, B.~Prunet-Foch and M.~Vignes-Adler, \emph{Physics of Fluids}, 2002,
  \textbf{14}, 3485--3501\relax
\mciteBstWouldAddEndPuncttrue
\mciteSetBstMidEndSepPunct{\mcitedefaultmidpunct}
{\mcitedefaultendpunct}{\mcitedefaultseppunct}\relax
\EndOfBibitem
\bibitem[Jha \emph{et~al.}(2020)Jha, Chantelot, Clanet, and
  Qu{\'e}r{\'e}]{jha2020viscous}
A.~Jha, P.~Chantelot, C.~Clanet and D.~Qu{\'e}r{\'e}, \emph{Soft Matter}, 2020,
  \textbf{16}, 7270--7273\relax
\mciteBstWouldAddEndPuncttrue
\mciteSetBstMidEndSepPunct{\mcitedefaultmidpunct}
{\mcitedefaultendpunct}{\mcitedefaultseppunct}\relax
\EndOfBibitem
\bibitem[Khojasteh \emph{et~al.}(2016)Khojasteh, Kazerooni, Salarian, and
  Kamali]{khojasteh2016droplet}
D.~Khojasteh, M.~Kazerooni, S.~Salarian and R.~Kamali, \emph{Journal of
  Industrial and Engineering Chemistry}, 2016, \textbf{42}, 1--14\relax
\mciteBstWouldAddEndPuncttrue
\mciteSetBstMidEndSepPunct{\mcitedefaultmidpunct}
{\mcitedefaultendpunct}{\mcitedefaultseppunct}\relax
\EndOfBibitem
\bibitem[Stow and Hadfield(1981)]{stow1981experimental}
C.~D. Stow and M.~G. Hadfield, \emph{Proceedings of the Royal Society of
  London. A. Mathematical and Physical Sciences}, 1981, \textbf{373},
  419--441\relax
\mciteBstWouldAddEndPuncttrue
\mciteSetBstMidEndSepPunct{\mcitedefaultmidpunct}
{\mcitedefaultendpunct}{\mcitedefaultseppunct}\relax
\EndOfBibitem
\bibitem[Mundo \emph{et~al.}(1995)Mundo, Sommerfeld, and Tropea]{MUNDO1995151}
C.~Mundo, M.~Sommerfeld and C.~Tropea, \emph{International Journal of
  Multiphase Flow}, 1995, \textbf{21}, 151 -- 173\relax
\mciteBstWouldAddEndPuncttrue
\mciteSetBstMidEndSepPunct{\mcitedefaultmidpunct}
{\mcitedefaultendpunct}{\mcitedefaultseppunct}\relax
\EndOfBibitem
\bibitem[Range and Feuillebois(1998)]{RANGE1998}
K.~Range and F.~Feuillebois, \emph{Journal of Colloid and Interface Science},
  1998, \textbf{203}, 16 -- 30\relax
\mciteBstWouldAddEndPuncttrue
\mciteSetBstMidEndSepPunct{\mcitedefaultmidpunct}
{\mcitedefaultendpunct}{\mcitedefaultseppunct}\relax
\EndOfBibitem
\bibitem[Latka \emph{et~al.}(2018)Latka, Boelens, Nagel, and
  de~Pablo]{latka2018drop}
A.~Latka, A.~M. Boelens, S.~R. Nagel and J.~J. de~Pablo, \emph{Physics of
  Fluids}, 2018, \textbf{30}, 022105\relax
\mciteBstWouldAddEndPuncttrue
\mciteSetBstMidEndSepPunct{\mcitedefaultmidpunct}
{\mcitedefaultendpunct}{\mcitedefaultseppunct}\relax
\EndOfBibitem
\bibitem[Quetzeri-Santiago \emph{et~al.}(2019)Quetzeri-Santiago, Yokoi,
  Castrej\'on-Pita, and Castrej\'on-Pita]{Quetzeri2019}
M.~A. Quetzeri-Santiago, K.~Yokoi, A.~A. Castrej\'on-Pita and J.~R.
  Castrej\'on-Pita, \emph{Phys. Rev. Lett.}, 2019, \textbf{122}, 228001\relax
\mciteBstWouldAddEndPuncttrue
\mciteSetBstMidEndSepPunct{\mcitedefaultmidpunct}
{\mcitedefaultendpunct}{\mcitedefaultseppunct}\relax
\EndOfBibitem
\bibitem[Blake and Ruschak(1979)]{blake1979maximum}
T.~D. Blake and K.~J. Ruschak, \emph{Nature}, 1979, \textbf{282},
  489--491\relax
\mciteBstWouldAddEndPuncttrue
\mciteSetBstMidEndSepPunct{\mcitedefaultmidpunct}
{\mcitedefaultendpunct}{\mcitedefaultseppunct}\relax
\EndOfBibitem
\bibitem[Snoeijer and Andreotti(2013)]{snoeijer2013}
J.~H. Snoeijer and B.~Andreotti, \emph{Annual review of fluid mechanics}, 2013,
  \textbf{45}, 269--292\relax
\mciteBstWouldAddEndPuncttrue
\mciteSetBstMidEndSepPunct{\mcitedefaultmidpunct}
{\mcitedefaultendpunct}{\mcitedefaultseppunct}\relax
\EndOfBibitem
\bibitem[Blake \emph{et~al.}(2023)Blake, Fern{\'a}ndez-Toledano, and
  De~Coninck]{blake2023possible}
T.~Blake, J.~Fern{\'a}ndez-Toledano and J.~De~Coninck, \emph{Journal of Colloid
  and Interface Science}, 2023, \textbf{629}, 660--669\relax
\mciteBstWouldAddEndPuncttrue
\mciteSetBstMidEndSepPunct{\mcitedefaultmidpunct}
{\mcitedefaultendpunct}{\mcitedefaultseppunct}\relax
\EndOfBibitem
\bibitem[Riboux and Gordillo(2017)]{Riboux2017}
G.~Riboux and J.~M. Gordillo, \emph{Phys. Rev. E}, 2017, \textbf{96},
  013105\relax
\mciteBstWouldAddEndPuncttrue
\mciteSetBstMidEndSepPunct{\mcitedefaultmidpunct}
{\mcitedefaultendpunct}{\mcitedefaultseppunct}\relax
\EndOfBibitem
\bibitem[de~Goede \emph{et~al.}(2018)de~Goede, Laan, de~Bruin, and
  Bonn]{deGoede2018}
T.~C. de~Goede, N.~Laan, K.~G. de~Bruin and D.~Bonn, \emph{Langmuir}, 2018,
  \textbf{34}, 5163--5168\relax
\mciteBstWouldAddEndPuncttrue
\mciteSetBstMidEndSepPunct{\mcitedefaultmidpunct}
{\mcitedefaultendpunct}{\mcitedefaultseppunct}\relax
\EndOfBibitem
\bibitem[Jian \emph{et~al.}(2018)Jian, Josserand, Popinet, Ray, and
  Zaleski]{jian2018}
Z.~Jian, C.~Josserand, S.~Popinet, P.~Ray and S.~Zaleski, \emph{Journal of
  Fluid Mechanics}, 2018, \textbf{835}, 1065--1086\relax
\mciteBstWouldAddEndPuncttrue
\mciteSetBstMidEndSepPunct{\mcitedefaultmidpunct}
{\mcitedefaultendpunct}{\mcitedefaultseppunct}\relax
\EndOfBibitem
\bibitem[Kolinski \emph{et~al.}(2019)Kolinski, Kaviani, Hade, and
  Rubinstein]{Kolinski2019}
J.~M. Kolinski, R.~Kaviani, D.~Hade and S.~M. Rubinstein, \emph{Phys. Rev.
  Fluids}, 2019, \textbf{4}, 123605\relax
\mciteBstWouldAddEndPuncttrue
\mciteSetBstMidEndSepPunct{\mcitedefaultmidpunct}
{\mcitedefaultendpunct}{\mcitedefaultseppunct}\relax
\EndOfBibitem
\bibitem[THORODDSEN \emph{et~al.}(2003)THORODDSEN, ETOH, and
  TAKEHARA]{Thoroddsen2003}
S.~T. THORODDSEN, T.~G. ETOH and K.~TAKEHARA, \emph{Journal of Fluid
  Mechanics}, 2003, \textbf{478}, 125–134\relax
\mciteBstWouldAddEndPuncttrue
\mciteSetBstMidEndSepPunct{\mcitedefaultmidpunct}
{\mcitedefaultendpunct}{\mcitedefaultseppunct}\relax
\EndOfBibitem
\bibitem[Driscoll and Nagel(2011)]{Driscoll2011}
M.~M. Driscoll and S.~R. Nagel, \emph{Phys. Rev. Lett.}, 2011, \textbf{107},
  154502\relax
\mciteBstWouldAddEndPuncttrue
\mciteSetBstMidEndSepPunct{\mcitedefaultmidpunct}
{\mcitedefaultendpunct}{\mcitedefaultseppunct}\relax
\EndOfBibitem
\bibitem[Liu \emph{et~al.}(2015)Liu, Tan, and Xu]{Liu2015}
Y.~Liu, P.~Tan and L.~Xu, \emph{Proceedings of the National Academy of
  Sciences}, 2015, \textbf{112}, 3280--3284\relax
\mciteBstWouldAddEndPuncttrue
\mciteSetBstMidEndSepPunct{\mcitedefaultmidpunct}
{\mcitedefaultendpunct}{\mcitedefaultseppunct}\relax
\EndOfBibitem
\bibitem[van~der Veen \emph{et~al.}(2012)van~der Veen, Tran, Lohse, and
  Sun]{vanderVeen2012}
R.~C.~A. van~der Veen, T.~Tran, D.~Lohse and C.~Sun, \emph{Phys. Rev. E}, 2012,
  \textbf{85}, 026315\relax
\mciteBstWouldAddEndPuncttrue
\mciteSetBstMidEndSepPunct{\mcitedefaultmidpunct}
{\mcitedefaultendpunct}{\mcitedefaultseppunct}\relax
\EndOfBibitem
\bibitem[Bird \emph{et~al.}(2009)Bird, Tsai, and Stone]{bird2009}
J.~C. Bird, S.~S. Tsai and H.~A. Stone, \emph{New Journal of Physics}, 2009,
  \textbf{11}, 063017\relax
\mciteBstWouldAddEndPuncttrue
\mciteSetBstMidEndSepPunct{\mcitedefaultmidpunct}
{\mcitedefaultendpunct}{\mcitedefaultseppunct}\relax
\EndOfBibitem
\bibitem[Hao \emph{et~al.}(2019)Hao, Lu, Lee, Wu, Hu, and Floryan]{Hao2019}
J.~Hao, J.~Lu, L.~Lee, Z.~Wu, G.~Hu and J.~Floryan, \emph{Physical review
  letters}, 2019, \textbf{122}, 054501\relax
\mciteBstWouldAddEndPuncttrue
\mciteSetBstMidEndSepPunct{\mcitedefaultmidpunct}
{\mcitedefaultendpunct}{\mcitedefaultseppunct}\relax
\EndOfBibitem
\bibitem[Garc{\'\i}a-Geijo \emph{et~al.}(2020)Garc{\'\i}a-Geijo, Riboux, and
  Gordillo]{Garcia2020}
P.~Garc{\'\i}a-Geijo, G.~Riboux and J.~M. Gordillo, \emph{Journal of Fluid
  Mechanics}, 2020, \textbf{897}, A12\relax
\mciteBstWouldAddEndPuncttrue
\mciteSetBstMidEndSepPunct{\mcitedefaultmidpunct}
{\mcitedefaultendpunct}{\mcitedefaultseppunct}\relax
\EndOfBibitem
\bibitem[Howland \emph{et~al.}(2016)Howland, Antkowiak, Castrej\'on-Pita,
  Howison, Oliver, Style, and Castrej\'on-Pita]{SoftSplash}
C.~J. Howland, A.~Antkowiak, J.~R. Castrej\'on-Pita, S.~D. Howison, J.~M.
  Oliver, R.~W. Style and A.~A. Castrej\'on-Pita, \emph{Phys. Rev. Lett.},
  2016, \textbf{117}, 184502\relax
\mciteBstWouldAddEndPuncttrue
\mciteSetBstMidEndSepPunct{\mcitedefaultmidpunct}
{\mcitedefaultendpunct}{\mcitedefaultseppunct}\relax
\EndOfBibitem
\bibitem[Ferziger and Peri\'{c}(2002)]{ferziger02:CMFD}
J.~H. Ferziger and M.~Peri\'{c}, \emph{{Computational Methods for Fluid
  Dynamics}}, Springer, Berlin, 3rd edn., 2002\relax
\mciteBstWouldAddEndPuncttrue
\mciteSetBstMidEndSepPunct{\mcitedefaultmidpunct}
{\mcitedefaultendpunct}{\mcitedefaultseppunct}\relax
\EndOfBibitem
\bibitem[Hughes(2012)]{hughes2012finite}
T.~J. Hughes, \emph{The finite element method: linear static and dynamic finite
  element analysis}, Courier Corporation, 2012\relax
\mciteBstWouldAddEndPuncttrue
\mciteSetBstMidEndSepPunct{\mcitedefaultmidpunct}
{\mcitedefaultendpunct}{\mcitedefaultseppunct}\relax
\EndOfBibitem
\bibitem[Anderson \emph{et~al.}(2020)Anderson, Tannehill, Pletcher, Munipalli,
  and Shankar]{anderson2020computational}
D.~Anderson, J.~C. Tannehill, R.~H. Pletcher, R.~Munipalli and V.~Shankar,
  \emph{Computational fluid mechanics and heat transfer}, CRC press, 2020\relax
\mciteBstWouldAddEndPuncttrue
\mciteSetBstMidEndSepPunct{\mcitedefaultmidpunct}
{\mcitedefaultendpunct}{\mcitedefaultseppunct}\relax
\EndOfBibitem
\bibitem[Kr{\"u}ger \emph{et~al.}(2017)Kr{\"u}ger, Kusumaatmaja, Kuzmin,
  Shardt, Silva, and Viggen]{kruger2017lattice}
T.~Kr{\"u}ger, H.~Kusumaatmaja, A.~Kuzmin, O.~Shardt, G.~Silva and E.~M.
  Viggen, \emph{Springer International Publishing}, 2017, \textbf{10},
  4--15\relax
\mciteBstWouldAddEndPuncttrue
\mciteSetBstMidEndSepPunct{\mcitedefaultmidpunct}
{\mcitedefaultendpunct}{\mcitedefaultseppunct}\relax
\EndOfBibitem
\bibitem[Glowinski and Pironneau(1992)]{glowinski1992finite}
R.~Glowinski and O.~Pironneau, \emph{Annual review of fluid mechanics}, 1992,
  \textbf{24}, 167--204\relax
\mciteBstWouldAddEndPuncttrue
\mciteSetBstMidEndSepPunct{\mcitedefaultmidpunct}
{\mcitedefaultendpunct}{\mcitedefaultseppunct}\relax
\EndOfBibitem
\bibitem[Kieckhefen \emph{et~al.}(2020)Kieckhefen, Pietsch, Dosta, and
  Heinrich]{kieckhefen2020possibilities}
P.~Kieckhefen, S.~Pietsch, M.~Dosta and S.~Heinrich, \emph{Annual review of
  chemical and biomolecular engineering}, 2020, \textbf{11}, 397--422\relax
\mciteBstWouldAddEndPuncttrue
\mciteSetBstMidEndSepPunct{\mcitedefaultmidpunct}
{\mcitedefaultendpunct}{\mcitedefaultseppunct}\relax
\EndOfBibitem
\bibitem[Gueyffier \emph{et~al.}(1999)Gueyffier, Li, Nadim, Scardovelli, and
  Zaleski]{gueyffier1999volume}
D.~Gueyffier, J.~Li, A.~Nadim, R.~Scardovelli and S.~Zaleski, \emph{Journal of
  Computational physics}, 1999, \textbf{152}, 423--456\relax
\mciteBstWouldAddEndPuncttrue
\mciteSetBstMidEndSepPunct{\mcitedefaultmidpunct}
{\mcitedefaultendpunct}{\mcitedefaultseppunct}\relax
\EndOfBibitem
\bibitem[Kim(2012)]{Kim_2012}
J.~Kim, \emph{Communications in Computational Physics}, 2012, \textbf{12},
  613–661\relax
\mciteBstWouldAddEndPuncttrue
\mciteSetBstMidEndSepPunct{\mcitedefaultmidpunct}
{\mcitedefaultendpunct}{\mcitedefaultseppunct}\relax
\EndOfBibitem
\bibitem[Sussman \emph{et~al.}(1994)Sussman, Smereka, and
  Osher]{sussman1994level}
M.~Sussman, P.~Smereka and S.~Osher, \emph{Journal of Computational physics},
  1994, \textbf{114}, 146--159\relax
\mciteBstWouldAddEndPuncttrue
\mciteSetBstMidEndSepPunct{\mcitedefaultmidpunct}
{\mcitedefaultendpunct}{\mcitedefaultseppunct}\relax
\EndOfBibitem
\bibitem[Osher and Fedkiw(2001)]{osher2001level}
S.~Osher and R.~P. Fedkiw, \emph{Journal of Computational physics}, 2001,
  \textbf{169}, 463--502\relax
\mciteBstWouldAddEndPuncttrue
\mciteSetBstMidEndSepPunct{\mcitedefaultmidpunct}
{\mcitedefaultendpunct}{\mcitedefaultseppunct}\relax
\EndOfBibitem
\bibitem[Osher and Sethian(1988)]{osher1988fronts}
S.~Osher and J.~A. Sethian, \emph{Journal of computational physics}, 1988,
  \textbf{79}, 12--49\relax
\mciteBstWouldAddEndPuncttrue
\mciteSetBstMidEndSepPunct{\mcitedefaultmidpunct}
{\mcitedefaultendpunct}{\mcitedefaultseppunct}\relax
\EndOfBibitem
\bibitem[Gunstensen \emph{et~al.}(1991)Gunstensen, Rothman, Zaleski, and
  Zanetti]{gunstensen1991lattice}
A.~K. Gunstensen, D.~H. Rothman, S.~Zaleski and G.~Zanetti, \emph{Physical
  review A}, 1991, \textbf{43}, 4320\relax
\mciteBstWouldAddEndPuncttrue
\mciteSetBstMidEndSepPunct{\mcitedefaultmidpunct}
{\mcitedefaultendpunct}{\mcitedefaultseppunct}\relax
\EndOfBibitem
\bibitem[Aidun and Clausen(2010)]{aidun2010lattice}
C.~K. Aidun and J.~R. Clausen, \emph{Annual review of fluid mechanics}, 2010,
  \textbf{42}, 439--472\relax
\mciteBstWouldAddEndPuncttrue
\mciteSetBstMidEndSepPunct{\mcitedefaultmidpunct}
{\mcitedefaultendpunct}{\mcitedefaultseppunct}\relax
\EndOfBibitem
\bibitem[Chen and Doolen(1998)]{chen1998lattice}
S.~Chen and G.~D. Doolen, \emph{Annual review of fluid mechanics}, 1998,
  \textbf{30}, 329--364\relax
\mciteBstWouldAddEndPuncttrue
\mciteSetBstMidEndSepPunct{\mcitedefaultmidpunct}
{\mcitedefaultendpunct}{\mcitedefaultseppunct}\relax
\EndOfBibitem
\bibitem[Pozorski and Olejnik(2023)]{pozorski2023smoothed}
J.~Pozorski and M.~Olejnik, \emph{Acta Mechanica}, 2023,  1--30\relax
\mciteBstWouldAddEndPuncttrue
\mciteSetBstMidEndSepPunct{\mcitedefaultmidpunct}
{\mcitedefaultendpunct}{\mcitedefaultseppunct}\relax
\EndOfBibitem
\bibitem[Wang \emph{et~al.}(2016)Wang, Chen, Wang, Liao, Zhu, and
  Li]{wang2016overview}
Z.-B. Wang, R.~Chen, H.~Wang, Q.~Liao, X.~Zhu and S.-Z. Li, \emph{Applied
  Mathematical Modelling}, 2016, \textbf{40}, 9625--9655\relax
\mciteBstWouldAddEndPuncttrue
\mciteSetBstMidEndSepPunct{\mcitedefaultmidpunct}
{\mcitedefaultendpunct}{\mcitedefaultseppunct}\relax
\EndOfBibitem
\bibitem[Bergeron \emph{et~al.}(2000)Bergeron, Bonn, Martin, and
  Vovelle]{Bergeron2000}
V.~Bergeron, D.~Bonn, J.~Y. Martin and L.~Vovelle, \emph{Nature}, 2000,
  \textbf{405}, 772--775\relax
\mciteBstWouldAddEndPuncttrue
\mciteSetBstMidEndSepPunct{\mcitedefaultmidpunct}
{\mcitedefaultendpunct}{\mcitedefaultseppunct}\relax
\EndOfBibitem
\bibitem[Crooks \emph{et~al.}(2001)Crooks, Cooper-White, and
  Boger]{CrooksCooperWhite2001}
R.~Crooks, J.~Cooper-White and D.~V. Boger, \emph{Chemical Engineering
  Science}, 2001, \textbf{56}, 5575 -- 5592\relax
\mciteBstWouldAddEndPuncttrue
\mciteSetBstMidEndSepPunct{\mcitedefaultmidpunct}
{\mcitedefaultendpunct}{\mcitedefaultseppunct}\relax
\EndOfBibitem
\bibitem[Bartolo \emph{et~al.}(2007)Bartolo, Boudaoud, Narcy, and
  Bonn]{Bartolo2007}
D.~Bartolo, A.~Boudaoud, G.~Narcy and D.~Bonn, \emph{Phys. Rev. Lett.}, 2007,
  \textbf{99}, 174502\relax
\mciteBstWouldAddEndPuncttrue
\mciteSetBstMidEndSepPunct{\mcitedefaultmidpunct}
{\mcitedefaultendpunct}{\mcitedefaultseppunct}\relax
\EndOfBibitem
\bibitem[Smith and Bertola(2010)]{SmithBertola2010}
M.~I. Smith and V.~Bertola, \emph{Phys. Rev. Lett.}, 2010, \textbf{104},
  154502\relax
\mciteBstWouldAddEndPuncttrue
\mciteSetBstMidEndSepPunct{\mcitedefaultmidpunct}
{\mcitedefaultendpunct}{\mcitedefaultseppunct}\relax
\EndOfBibitem
\bibitem[Zang \emph{et~al.}(2013)Zang, Wang, Geng, Zhang, and Chen]{Zang2013}
D.~Zang, X.~Wang, X.~Geng, Y.~Zhang and Y.~Chen, \emph{Soft Matter}, 2013,
  \textbf{9}, 394--400\relax
\mciteBstWouldAddEndPuncttrue
\mciteSetBstMidEndSepPunct{\mcitedefaultmidpunct}
{\mcitedefaultendpunct}{\mcitedefaultseppunct}\relax
\EndOfBibitem
\bibitem[Crooks and Boger(2000)]{CrooksBoger2000}
R.~Crooks and D.~V. Boger, \emph{Journal of Rheology}, 2000, \textbf{44},
  973--996\relax
\mciteBstWouldAddEndPuncttrue
\mciteSetBstMidEndSepPunct{\mcitedefaultmidpunct}
{\mcitedefaultendpunct}{\mcitedefaultseppunct}\relax
\EndOfBibitem
\bibitem[Huh \emph{et~al.}(2015)Huh, Jung, Seo, and Lee]{huh2015role}
H.~K. Huh, S.~Jung, K.~W. Seo and S.~J. Lee, \emph{Microfluidics and
  Nanofluidics}, 2015, \textbf{18}, 1221--1232\relax
\mciteBstWouldAddEndPuncttrue
\mciteSetBstMidEndSepPunct{\mcitedefaultmidpunct}
{\mcitedefaultendpunct}{\mcitedefaultseppunct}\relax
\EndOfBibitem
\bibitem[Rahimi and Weihs(2011)]{GelledFuel}
S.~Rahimi and D.~Weihs, \emph{Propellants, Explosives, Pyrotechnics}, 2011,
  \textbf{36}, 273--281\relax
\mciteBstWouldAddEndPuncttrue
\mciteSetBstMidEndSepPunct{\mcitedefaultmidpunct}
{\mcitedefaultendpunct}{\mcitedefaultseppunct}\relax
\EndOfBibitem
\bibitem[Bertola(2013)]{BERTOLA2013polymerimpact}
V.~Bertola, \emph{Advances in Colloid and Interface Science}, 2013,
  \textbf{193-194}, 1--11\relax
\mciteBstWouldAddEndPuncttrue
\mciteSetBstMidEndSepPunct{\mcitedefaultmidpunct}
{\mcitedefaultendpunct}{\mcitedefaultseppunct}\relax
\EndOfBibitem
\bibitem[An and Lee(2012)]{AnLee2012}
S.~M. An and S.~Y. Lee, \emph{Experimental Thermal and Fluid Science}, 2012,
  \textbf{38}, 140 -- 148\relax
\mciteBstWouldAddEndPuncttrue
\mciteSetBstMidEndSepPunct{\mcitedefaultmidpunct}
{\mcitedefaultendpunct}{\mcitedefaultseppunct}\relax
\EndOfBibitem
\bibitem[Scheller and Bousfield(1995)]{SchellerBousfieldNewtonian}
B.~L. Scheller and D.~W. Bousfield, \emph{AIChE Journal}, 1995, \textbf{41},
  1357--1367\relax
\mciteBstWouldAddEndPuncttrue
\mciteSetBstMidEndSepPunct{\mcitedefaultmidpunct}
{\mcitedefaultendpunct}{\mcitedefaultseppunct}\relax
\EndOfBibitem
\bibitem[Rozhkov \emph{et~al.}(2003)Rozhkov, Prunet-Foch, and
  Vignes-Adler]{Rozhkov2003}
A.~Rozhkov, B.~Prunet-Foch and M.~Vignes-Adler, \emph{Physics of Fluids}, 2003,
  \textbf{15}, 2006--2019\relax
\mciteBstWouldAddEndPuncttrue
\mciteSetBstMidEndSepPunct{\mcitedefaultmidpunct}
{\mcitedefaultendpunct}{\mcitedefaultseppunct}\relax
\EndOfBibitem
\bibitem[Lee \emph{et~al.}(2023)Lee, Kim, and Choi]{lee2023rational}
S.~J. Lee, K.~Kim and W.~Choi, \emph{Applied Physics Letters}, 2023,
  \textbf{122}, 261601\relax
\mciteBstWouldAddEndPuncttrue
\mciteSetBstMidEndSepPunct{\mcitedefaultmidpunct}
{\mcitedefaultendpunct}{\mcitedefaultseppunct}\relax
\EndOfBibitem
\bibitem[Luu and Forterre(2009)]{luu_forterre_2009}
L.-H. Luu and Y.~Forterre, \emph{Journal of Fluid Mechanics}, 2009,
  \textbf{632}, 301–327\relax
\mciteBstWouldAddEndPuncttrue
\mciteSetBstMidEndSepPunct{\mcitedefaultmidpunct}
{\mcitedefaultendpunct}{\mcitedefaultseppunct}\relax
\EndOfBibitem
\bibitem[Luu and Forterre(2013)]{LuuForterre2013}
L.-H. Luu and Y.~Forterre, \emph{Phys. Rev. Lett.}, 2013, \textbf{110},
  184501\relax
\mciteBstWouldAddEndPuncttrue
\mciteSetBstMidEndSepPunct{\mcitedefaultmidpunct}
{\mcitedefaultendpunct}{\mcitedefaultseppunct}\relax
\EndOfBibitem
\bibitem[Gu\'emas \emph{et~al.}(2012)Gu\'emas, Mar\'in, and Lohse]{Guemas}
M.~Gu\'emas, A.~G. Mar\'in and D.~Lohse, \emph{Soft Matter}, 2012, \textbf{8},
  10725--10731\relax
\mciteBstWouldAddEndPuncttrue
\mciteSetBstMidEndSepPunct{\mcitedefaultmidpunct}
{\mcitedefaultendpunct}{\mcitedefaultseppunct}\relax
\EndOfBibitem
\bibitem[Mohammad~Karim(2019)]{mohammad2019experimental}
A.~Mohammad~Karim, \emph{AIP Advances}, 2019, \textbf{9}, 125141\relax
\mciteBstWouldAddEndPuncttrue
\mciteSetBstMidEndSepPunct{\mcitedefaultmidpunct}
{\mcitedefaultendpunct}{\mcitedefaultseppunct}\relax
\EndOfBibitem
\bibitem[Cheng \emph{et~al.}(2022)Cheng, Sun, and Gordillo]{ChengStressReview}
X.~Cheng, T.-P. Sun and L.~Gordillo, \emph{Annual Review of Fluid Mechanics},
  2022, \textbf{54}, 57--81\relax
\mciteBstWouldAddEndPuncttrue
\mciteSetBstMidEndSepPunct{\mcitedefaultmidpunct}
{\mcitedefaultendpunct}{\mcitedefaultseppunct}\relax
\EndOfBibitem
\bibitem[Kim and Baek(2012)]{KIM201262}
E.~Kim and J.~Baek, \emph{Journal of Non-Newtonian Fluid Mechanics}, 2012,
  \textbf{173-174}, 62 -- 71\relax
\mciteBstWouldAddEndPuncttrue
\mciteSetBstMidEndSepPunct{\mcitedefaultmidpunct}
{\mcitedefaultendpunct}{\mcitedefaultseppunct}\relax
\EndOfBibitem
\bibitem[Oishi \emph{et~al.}(2019)Oishi, Thompson, and
  Martins]{oishi_thompson_martins_2019}
C.~M. Oishi, R.~L. Thompson and F.~P. Martins, \emph{Journal of Fluid
  Mechanics}, 2019, \textbf{876}, 642–679\relax
\mciteBstWouldAddEndPuncttrue
\mciteSetBstMidEndSepPunct{\mcitedefaultmidpunct}
{\mcitedefaultendpunct}{\mcitedefaultseppunct}\relax
\EndOfBibitem
\bibitem[Morris(2020)]{morris2020toward}
J.~F. Morris, \emph{Physical Review Fluids}, 2020, \textbf{5}, 110519\relax
\mciteBstWouldAddEndPuncttrue
\mciteSetBstMidEndSepPunct{\mcitedefaultmidpunct}
{\mcitedefaultendpunct}{\mcitedefaultseppunct}\relax
\EndOfBibitem
\bibitem[Ness \emph{et~al.}(2022)Ness, Seto, and Mari]{ness2022physics}
C.~Ness, R.~Seto and R.~Mari, \emph{Annual Review of Condensed Matter Physics},
  2022, \textbf{13}, 97--117\relax
\mciteBstWouldAddEndPuncttrue
\mciteSetBstMidEndSepPunct{\mcitedefaultmidpunct}
{\mcitedefaultendpunct}{\mcitedefaultseppunct}\relax
\EndOfBibitem
\bibitem[German and Bertola(2009)]{German_2009}
G.~German and V.~Bertola, \emph{Journal of Physics: Condensed Matter}, 2009,
  \textbf{21}, 375111\relax
\mciteBstWouldAddEndPuncttrue
\mciteSetBstMidEndSepPunct{\mcitedefaultmidpunct}
{\mcitedefaultendpunct}{\mcitedefaultseppunct}\relax
\EndOfBibitem
\bibitem[Jørgensen \emph{et~al.}(2020)Jørgensen, Forterre, and
  Lhuissier]{jorgensen2020deformation}
L.~Jørgensen, Y.~Forterre and H.~Lhuissier, \emph{Deformation upon impact of a
  concentrated suspension drop}, 2020\relax
\mciteBstWouldAddEndPuncttrue
\mciteSetBstMidEndSepPunct{\mcitedefaultmidpunct}
{\mcitedefaultendpunct}{\mcitedefaultseppunct}\relax
\EndOfBibitem
\bibitem[Ok \emph{et~al.}(2005)Ok, Park, Carr, Morris, and Zhu]{Ok}
H.~Ok, H.~Park, W.~W. Carr, J.~F. Morris and J.~Zhu, \emph{Journal of
  Dispersion Science and Technology}, 2005, \textbf{25}, 449--456\relax
\mciteBstWouldAddEndPuncttrue
\mciteSetBstMidEndSepPunct{\mcitedefaultmidpunct}
{\mcitedefaultendpunct}{\mcitedefaultseppunct}\relax
\EndOfBibitem
\bibitem[NICOLAS(2005)]{nicolas_2005}
M.~NICOLAS, \emph{Journal of Fluid Mechanics}, 2005, \textbf{545},
  271–280\relax
\mciteBstWouldAddEndPuncttrue
\mciteSetBstMidEndSepPunct{\mcitedefaultmidpunct}
{\mcitedefaultendpunct}{\mcitedefaultseppunct}\relax
\EndOfBibitem
\bibitem[Boyer \emph{et~al.}(2016)Boyer, Sandoval-Nava, Snoeijer, Dijksman, and
  Lohse]{Boyer}
F.~m.~c. Boyer, E.~Sandoval-Nava, J.~H. Snoeijer, J.~F. Dijksman and D.~Lohse,
  \emph{Phys. Rev. Fluids}, 2016, \textbf{1}, 013901\relax
\mciteBstWouldAddEndPuncttrue
\mciteSetBstMidEndSepPunct{\mcitedefaultmidpunct}
{\mcitedefaultendpunct}{\mcitedefaultseppunct}\relax
\EndOfBibitem
\bibitem[Raux \emph{et~al.}(2020)Raux, Troger, Jop, and Sauret]{Raux}
P.~S. Raux, A.~Troger, P.~Jop and A.~Sauret, \emph{Phys. Rev. Fluids}, 2020,
  \textbf{5}, 044004\relax
\mciteBstWouldAddEndPuncttrue
\mciteSetBstMidEndSepPunct{\mcitedefaultmidpunct}
{\mcitedefaultendpunct}{\mcitedefaultseppunct}\relax
\EndOfBibitem
\bibitem[Grishaev \emph{et~al.}(2015)Grishaev, Iorio, Dubois, and
  Amirfazli]{Grishaev}
V.~Grishaev, C.~S. Iorio, F.~Dubois and A.~Amirfazli, \emph{Langmuir}, 2015,
  \textbf{31}, 9833--9844\relax
\mciteBstWouldAddEndPuncttrue
\mciteSetBstMidEndSepPunct{\mcitedefaultmidpunct}
{\mcitedefaultendpunct}{\mcitedefaultseppunct}\relax
\EndOfBibitem
\bibitem[Grishaev \emph{et~al.}(2017)Grishaev, Iorio, Dubois, and
  Amirfazli]{GRISHAEVdist}
V.~Grishaev, C.~S. Iorio, F.~Dubois and A.~Amirfazli, \emph{Journal of Colloid
  and Interface Science}, 2017, \textbf{490}, 108--118\relax
\mciteBstWouldAddEndPuncttrue
\mciteSetBstMidEndSepPunct{\mcitedefaultmidpunct}
{\mcitedefaultendpunct}{\mcitedefaultseppunct}\relax
\EndOfBibitem
\bibitem[Bertola and Haw(2015)]{bertola2015impact}
V.~Bertola and M.~D. Haw, \emph{Powder Technology}, 2015, \textbf{270},
  412--417\relax
\mciteBstWouldAddEndPuncttrue
\mciteSetBstMidEndSepPunct{\mcitedefaultmidpunct}
{\mcitedefaultendpunct}{\mcitedefaultseppunct}\relax
\EndOfBibitem
\bibitem[Kim \emph{et~al.}(2019)Kim, Kim, Lee, and Jeon]{KIM2019QCM}
G.~Kim, W.~Kim, S.~Lee and S.~Jeon, \emph{Sensors and Actuators B: Chemical},
  2019, \textbf{288}, 716--720\relax
\mciteBstWouldAddEndPuncttrue
\mciteSetBstMidEndSepPunct{\mcitedefaultmidpunct}
{\mcitedefaultendpunct}{\mcitedefaultseppunct}\relax
\EndOfBibitem
\bibitem[Waitukaitis and Jaeger(2012)]{waitukaitis2012impact}
S.~R. Waitukaitis and H.~M. Jaeger, \emph{Nature}, 2012, \textbf{487},
  205--209\relax
\mciteBstWouldAddEndPuncttrue
\mciteSetBstMidEndSepPunct{\mcitedefaultmidpunct}
{\mcitedefaultendpunct}{\mcitedefaultseppunct}\relax
\EndOfBibitem
\bibitem[Han \emph{et~al.}(2016)Han, Peters, and Jaeger]{han2016high}
E.~Han, I.~R. Peters and H.~M. Jaeger, \emph{Nature communications}, 2016,
  \textbf{7}, 1--8\relax
\mciteBstWouldAddEndPuncttrue
\mciteSetBstMidEndSepPunct{\mcitedefaultmidpunct}
{\mcitedefaultendpunct}{\mcitedefaultseppunct}\relax
\EndOfBibitem
\bibitem[Peters \emph{et~al.}(2016)Peters, Majumdar, and
  Jaeger]{peters2016direct}
I.~R. Peters, S.~Majumdar and H.~M. Jaeger, \emph{Nature}, 2016, \textbf{532},
  214--217\relax
\mciteBstWouldAddEndPuncttrue
\mciteSetBstMidEndSepPunct{\mcitedefaultmidpunct}
{\mcitedefaultendpunct}{\mcitedefaultseppunct}\relax
\EndOfBibitem
\bibitem[Han \emph{et~al.}(2018)Han, Wyart, Peters, and
  Jaeger]{HanJaegerShearFronts}
E.~Han, M.~Wyart, I.~R. Peters and H.~M. Jaeger, \emph{Phys. Rev. Fluids},
  2018, \textbf{3}, 073301\relax
\mciteBstWouldAddEndPuncttrue
\mciteSetBstMidEndSepPunct{\mcitedefaultmidpunct}
{\mcitedefaultendpunct}{\mcitedefaultseppunct}\relax
\EndOfBibitem
\bibitem[R\o{}mcke \emph{et~al.}(2021)R\o{}mcke, Peters, and
  Hearst]{RomckePetersPRF2021}
O.~R\o{}mcke, I.~R. Peters and R.~J. Hearst, \emph{Phys. Rev. Fluids}, 2021,
  \textbf{6}, 063301\relax
\mciteBstWouldAddEndPuncttrue
\mciteSetBstMidEndSepPunct{\mcitedefaultmidpunct}
{\mcitedefaultendpunct}{\mcitedefaultseppunct}\relax
\EndOfBibitem
\bibitem[Philippi \emph{et~al.}(2016)Philippi, Lagrée, and
  Antkowiak]{philippi_2016_newtonian}
J.~Philippi, P.-Y. Lagrée and A.~Antkowiak, \emph{Journal of Fluid Mechanics},
  2016, \textbf{795}, 96–135\relax
\mciteBstWouldAddEndPuncttrue
\mciteSetBstMidEndSepPunct{\mcitedefaultmidpunct}
{\mcitedefaultendpunct}{\mcitedefaultseppunct}\relax
\EndOfBibitem
\bibitem[Gordillo \emph{et~al.}(2018)Gordillo, Sun, and
  Cheng]{gordillo_force_2018}
L.~Gordillo, T.-P. Sun and X.~Cheng, \emph{Journal of Fluid Mechanics}, 2018,
  \textbf{840}, 190–214\relax
\mciteBstWouldAddEndPuncttrue
\mciteSetBstMidEndSepPunct{\mcitedefaultmidpunct}
{\mcitedefaultendpunct}{\mcitedefaultseppunct}\relax
\EndOfBibitem
\bibitem[Style \emph{et~al.}(2014)Style, Boltyanskiy, German, Hyland, MacMinn,
  Mertz, Wilen, Xu, and Dufresne]{style2014traction}
R.~W. Style, R.~Boltyanskiy, G.~K. German, C.~Hyland, C.~W. MacMinn, A.~F.
  Mertz, L.~A. Wilen, Y.~Xu and E.~R. Dufresne, \emph{Soft matter}, 2014,
  \textbf{10}, 4047--4055\relax
\mciteBstWouldAddEndPuncttrue
\mciteSetBstMidEndSepPunct{\mcitedefaultmidpunct}
{\mcitedefaultendpunct}{\mcitedefaultseppunct}\relax
\EndOfBibitem
\bibitem[Rathee \emph{et~al.}(2020)Rathee, Blair, and Urbach]{Rathee_JRheol}
V.~Rathee, D.~L. Blair and J.~S. Urbach, \emph{Journal of Rheology}, 2020,
  \textbf{64}, 299--308\relax
\mciteBstWouldAddEndPuncttrue
\mciteSetBstMidEndSepPunct{\mcitedefaultmidpunct}
{\mcitedefaultendpunct}{\mcitedefaultseppunct}\relax
\EndOfBibitem
\bibitem[Rathee \emph{et~al.}(2022)Rathee, Miller, Blair, and
  Urbach]{rathee2022structure}
V.~Rathee, J.~M. Miller, D.~L. Blair and J.~S. Urbach, \emph{arXiv preprint
  arXiv:2203.02482}, 2022\relax
\mciteBstWouldAddEndPuncttrue
\mciteSetBstMidEndSepPunct{\mcitedefaultmidpunct}
{\mcitedefaultendpunct}{\mcitedefaultseppunct}\relax
\EndOfBibitem
\bibitem[Williams and Philipse(2003)]{williams2003}
S.~Williams and A.~Philipse, \emph{Physical Review E}, 2003, \textbf{67},
  051301\relax
\mciteBstWouldAddEndPuncttrue
\mciteSetBstMidEndSepPunct{\mcitedefaultmidpunct}
{\mcitedefaultendpunct}{\mcitedefaultseppunct}\relax
\EndOfBibitem
\bibitem[James \emph{et~al.}(2019)James, Xue, Goyal, and Jaeger]{james2019c}
N.~M. James, H.~Xue, M.~Goyal and H.~M. Jaeger, \emph{Soft matter}, 2019,
  \textbf{15}, 3649--3654\relax
\mciteBstWouldAddEndPuncttrue
\mciteSetBstMidEndSepPunct{\mcitedefaultmidpunct}
{\mcitedefaultendpunct}{\mcitedefaultseppunct}\relax
\EndOfBibitem
\bibitem[Peters \emph{et~al.}(2013)Peters, Xu, and Jaeger]{peters2013splashing}
I.~R. Peters, Q.~Xu and H.~M. Jaeger, \emph{Physical review letters}, 2013,
  \textbf{111}, 028301\relax
\mciteBstWouldAddEndPuncttrue
\mciteSetBstMidEndSepPunct{\mcitedefaultmidpunct}
{\mcitedefaultendpunct}{\mcitedefaultseppunct}\relax
\EndOfBibitem
\bibitem[Klein~Schaarsberg \emph{et~al.}(2016)Klein~Schaarsberg, Peters, Stern,
  Dodge, Zhang, and Jaeger]{Schaarsberg}
M.~H. Klein~Schaarsberg, I.~R. Peters, M.~Stern, K.~Dodge, W.~W. Zhang and
  H.~M. Jaeger, \emph{Phys. Rev. E}, 2016, \textbf{93}, 062609\relax
\mciteBstWouldAddEndPuncttrue
\mciteSetBstMidEndSepPunct{\mcitedefaultmidpunct}
{\mcitedefaultendpunct}{\mcitedefaultseppunct}\relax
\EndOfBibitem
\bibitem[Mitarai and Nori(2006)]{wetGranularReview}
N.~Mitarai and F.~Nori, \emph{Advances in Physics}, 2006, \textbf{55},
  1--45\relax
\mciteBstWouldAddEndPuncttrue
\mciteSetBstMidEndSepPunct{\mcitedefaultmidpunct}
{\mcitedefaultendpunct}{\mcitedefaultseppunct}\relax
\EndOfBibitem
\bibitem[Lubbers \emph{et~al.}(2014)Lubbers, Xu, Wilken, Zhang, and
  Jaeger]{Lubbers}
L.~A. Lubbers, Q.~Xu, S.~Wilken, W.~W. Zhang and H.~M. Jaeger, \emph{Phys. Rev.
  Lett.}, 2014, \textbf{113}, 044502\relax
\mciteBstWouldAddEndPuncttrue
\mciteSetBstMidEndSepPunct{\mcitedefaultmidpunct}
{\mcitedefaultendpunct}{\mcitedefaultseppunct}\relax
\EndOfBibitem
\bibitem[Peters \emph{et~al.}(2013)Peters, Xu, and Jaeger]{Peters}
I.~R. Peters, Q.~Xu and H.~M. Jaeger, \emph{Phys. Rev. Lett.}, 2013,
  \textbf{111}, 028301\relax
\mciteBstWouldAddEndPuncttrue
\mciteSetBstMidEndSepPunct{\mcitedefaultmidpunct}
{\mcitedefaultendpunct}{\mcitedefaultseppunct}\relax
\EndOfBibitem
\bibitem[Grishaev \emph{et~al.}(2019)Grishaev, Bakulin, and
  Akhatov]{GRISHAEV2019_splashing}
V.~G. Grishaev, I.~K. Bakulin and I.~S. Akhatov, \emph{Journal of Fluids and
  Structures}, 2019, \textbf{91}, 102718\relax
\mciteBstWouldAddEndPuncttrue
\mciteSetBstMidEndSepPunct{\mcitedefaultmidpunct}
{\mcitedefaultendpunct}{\mcitedefaultseppunct}\relax
\EndOfBibitem
\bibitem[Stickel and Powell(2005)]{StickelPowellRheo}
J.~J. Stickel and R.~L. Powell, \emph{Annual Review of Fluid Mechanics}, 2005,
  \textbf{37}, 129--149\relax
\mciteBstWouldAddEndPuncttrue
\mciteSetBstMidEndSepPunct{\mcitedefaultmidpunct}
{\mcitedefaultendpunct}{\mcitedefaultseppunct}\relax
\EndOfBibitem
\bibitem[Westerweel \emph{et~al.}(2013)Westerweel, Elsinga, and
  Adrian]{westerweel2013particle}
J.~Westerweel, G.~E. Elsinga and R.~J. Adrian, \emph{Annual Review of Fluid
  Mechanics}, 2013, \textbf{45}, 409--436\relax
\mciteBstWouldAddEndPuncttrue
\mciteSetBstMidEndSepPunct{\mcitedefaultmidpunct}
{\mcitedefaultendpunct}{\mcitedefaultseppunct}\relax
\EndOfBibitem
\bibitem[Tang \emph{et~al.}(2019)Tang, Saha, Sun, and Law]{tang2019spreading}
X.~Tang, A.~Saha, C.~Sun and C.~K. Law, \emph{Journal of Fluid Mechanics},
  2019, \textbf{881}, 859--871\relax
\mciteBstWouldAddEndPuncttrue
\mciteSetBstMidEndSepPunct{\mcitedefaultmidpunct}
{\mcitedefaultendpunct}{\mcitedefaultseppunct}\relax
\EndOfBibitem
\bibitem[Patil \emph{et~al.}(2016)Patil, Bange, Bhardwaj, and
  Sharma]{patil2016effects}
N.~D. Patil, P.~G. Bange, R.~Bhardwaj and A.~Sharma, \emph{Langmuir}, 2016,
  \textbf{32}, 11958--11972\relax
\mciteBstWouldAddEndPuncttrue
\mciteSetBstMidEndSepPunct{\mcitedefaultmidpunct}
{\mcitedefaultendpunct}{\mcitedefaultseppunct}\relax
\EndOfBibitem
\bibitem[Ma \emph{et~al.}(2024)Ma, Aldhaleai, Liu, and Tsai]{ma2024nanofluid}
X.~Ma, A.~Aldhaleai, L.~Liu and P.~A. Tsai, \emph{Langmuir}, 2024\relax
\mciteBstWouldAddEndPuncttrue
\mciteSetBstMidEndSepPunct{\mcitedefaultmidpunct}
{\mcitedefaultendpunct}{\mcitedefaultseppunct}\relax
\EndOfBibitem
\bibitem[Waitukaitis \emph{et~al.}(2017)Waitukaitis, Zuiderwijk, Souslov,
  Coulais, and Van~Hecke]{waitukaitis2017coupling}
S.~R. Waitukaitis, A.~Zuiderwijk, A.~Souslov, C.~Coulais and M.~Van~Hecke,
  \emph{Nature Physics}, 2017, \textbf{13}, 1095--1099\relax
\mciteBstWouldAddEndPuncttrue
\mciteSetBstMidEndSepPunct{\mcitedefaultmidpunct}
{\mcitedefaultendpunct}{\mcitedefaultseppunct}\relax
\EndOfBibitem
\bibitem[Pham \emph{et~al.}(2017)Pham, Paven, Wooh, Kajiya, Butt, and
  Vollmer]{pham2017spontaneous}
J.~T. Pham, M.~Paven, S.~Wooh, T.~Kajiya, H.-J. Butt and D.~Vollmer,
  \emph{Nature communications}, 2017, \textbf{8}, 905\relax
\mciteBstWouldAddEndPuncttrue
\mciteSetBstMidEndSepPunct{\mcitedefaultmidpunct}
{\mcitedefaultendpunct}{\mcitedefaultseppunct}\relax
\EndOfBibitem
\end{mcitethebibliography}
\bibliographystyle{rsc} 

\end{document}